\documentclass{elsarticle}
\usepackage[ruled]{algorithm2e}
\usepackage{times}
\usepackage{latexsym}
\usepackage{todonotes}
\usepackage{footmisc}
\usepackage{url}
\usepackage{mathtools}
\usepackage{float}
\usepackage{xcolor}
\usepackage{subcaption}
\usepackage{multirow}
\usepackage{atbegshi}
\usepackage{enumitem}
\usepackage{amsmath}
\usepackage{amssymb}
\usepackage{pifont}
\usepackage{booktabs}

\SetAlFnt{\small}
\SetAlCapFnt{\small}
\SetAlCapNameFnt{\small}
\SetAlCapHSkip{0pt}
\IncMargin{-\parindent}

\newcommand{\word}[1]{\textsf{#1}}              
\newcommand{\gist}[1]{\textsc{#1}}             
\newcommand{\wikipage}[1]{\textsc{#1}}             
\newcommand{\wikicategory}[1]{\textsc{#1}}       

\begin{document}

\markboth{L. Weiland et al.}{Knowledge-rich Image Gist Understanding Beyond Literal Meaning}


\title{Knowledge-rich Image Gist Understanding\\
Beyond Literal Meaning}
\author[uma]{Lydia Weiland}
\author[uma]{Ioana Hulpu\c{s}}
\author[uma]{Simone Paolo Ponzetto}
\author[uma]{Wolfgang Effelsberg}

\author[unh]{Laura Dietz}

\address[uma]{University of Mannheim, B6 26, 68159 Mannheim, Germany}
\address[unh]{University of New Hampshire, 105 Main St, Durham, New Hampshire 03824, USA}

\begin{abstract}
We investigate the problem of understanding the message (\emph{gist}) conveyed by images and their captions as found, for instance, on websites or news articles. To this end, we propose a methodology to capture the meaning of image-caption pairs on the basis of large amounts of machine-readable knowledge that has previously been shown to be highly effective for text understanding. 
Our method identifies the connotation of objects beyond their denotation: where most approaches to image understanding focus on the denotation of objects, i.e., their literal meaning, our work addresses the identification of connotations, i.e., iconic meanings of objects,  to understand the message of images. We view image understanding as the task of representing an image-caption pair on the basis of a wide-coverage vocabulary of concepts such as the one provided by Wikipedia, and cast gist detection as a concept-ranking problem with image-caption pairs as queries.
%
%
Our proposed algorithm brings together aspects of entity linking and clustering, subgraph selection, semantic relatedness, and learning-to-rank in a novel way. In addition to this novel task and a complete evaluation of our approach, we introduce a novel dataset to foster further research on this problem. To enable a thorough investigation of the problem of gist understanding, we produce a gold standard of over 300 image-caption pairs and over 8,000 gist annotations covering a wide variety of topics at different levels of abstraction.
We use this dataset to experimentally benchmark the contribution of signals from heterogeneous sources, namely image and text. The best result with a Mean Average Precision (MAP) of 0.69 indicate that by combining both dimensions we are able to better understand the meaning of our image-caption pairs than when using language or vision information alone. 
We test the robustness of our gist detection approach when receiving automatically generated input, i.e., using automatically generated image tags or generated captions, and prove the feasibility of an end-to-end automated process. However, we also show experimentally that state-of-the-art image and text understanding is better at dealing with literal meanings of image-caption pairs, with non-literal pairs being instead generally more difficult to detect, thus paving the way for future work on understanding the message of images beyond their literal content.
\end{abstract}

%
%


%
%






\maketitle

\section{Introduction}\label{introduction}

%
\noindent Newspaper articles and blog posts are often accompanied by figures, which consist of an image and a caption. While in some cases figures are used as mere decoration, more often figures support the message of the article in stimulating emotions and transmitting intentions \cite{LuxKM10,HanjalicKL12}. This is especially the case on matters of controversial topics, such as, for instance, global warming, where emotions are conveyed through so-called media icons~\cite{Perlmutter2004,Drechsel2010}: images with high suggestive power that illustrate the topic. A picture of a polar bear on melting shelf ice is a famous example cited by advocates stopping carbon emissions~\cite{oneill14}. As such, many image-caption pairs are able to broadcast abstract concepts and emotions \cite{oneill09} beyond the physical objects they illustrate.

%
Previous research in image understanding has focused on the identification and labeling of objects that are visible in the image (e.g., PascalVOC~\cite{Everingham2010}, MS COCO~\cite{502}, Im2Text~\cite{Ordonez2011}, to name a few, cf.\ also Section \ref{rw}). Recently, the captionbot system \cite{tran2016captionbot} was proposed to generate captions for a given image. However, all these approaches focus on the description of what is to be explicitly found, i.e., \emph{depictable}, within pictures.\footnote{Throughout the paper, we use the terms \emph{depictable} and \emph{non-depictable} to refer to concrete and abstract aspects of image-caption pairs and their gists, respectively.} For the example in Figure 1b, captionbot generates the caption: ``I think it's a brown bear sitting on a bench.'' But despite many research efforts having focused on the so-called problem of bridging the semantic gap in both automatic image and text analysis -- namely, the process of replacing low-level (visual and textual) descriptors with higher-level semantically-rich ones -- few papers looked at the complementary, even more challenging problem of bridging the intentional gap, namely understanding the intention behind using a specific image in context \cite{Kofler16}.

%
%
In this paper, we take a first step towards addressing this hard problem by presenting a method to identify the message that an image conveys, including also abstract (i.e., \emph{non-depictable}) topics. Specifically, we look at the task of identifying and ranking concepts that capture the message of the image, hereafter called \emph{gist}. Starting from the visible objects in the image and entity mentions in the caption, we study the use of external knowledge bases for the identification of concepts that represent the gist of the image.\footnote{Hereafter, we use \emph{concept} and \emph{entity} interchangeably to refer to resources of the knowledge base.} Thus, we cast the problem of gist detection as a concept ranking task with the following twist: Given an image-caption pair, rank concepts from Wikipedia according to how well they express the gist of the image-caption pair.

The contributions of this paper are the following:

\begin{itemize}[leftmargin=4mm]

\item \textbf{New task}: we formulate the novel task of detecting the gist of image-caption pairs using the vocabulary and topics provided by a reference external resource, i.e., a knowledge base.

\item \textbf{New framework}: we present a methodology to identify the gist of image-caption pairs using a supervised ranking model that combines content- and graph-based features from an underlying knowledge base, i.e., Wikipedia in our case. Our approach effectively combines object detection, text disambiguation, knowledge-based topic detection, and learning-to-rank in a novel way.

\item \textbf{Experimental study}: In order to promote a community around gist detection and foster further research, we create a gold standard that we use to investigate a wide range of research questions associated with our task.

\end{itemize}

\medskip\noindent\textbf{Outline.} In the remainder of this paper, we first review related work in Section \ref{rw} and move on to define the task of image gist detection in Section~\ref{problemStatement}. We present our approach in Section~\ref{approach} and then conduct an extensive evaluation in Section~\ref{experiments}, where we investigate ten different research questions related to our problem. We conclude with final remarks and future work directions in Section \ref{concl}.

\section{Related Work}\label{rw}

\noindent
To the best of our knowledge, this is the first work to view the task of image gist understanding as an entity ranking problem using the structure and vocabulary of a knowledge base. This complements very recent work in the field of computer vision on knowledge-aware object detection \cite{ijcai2017-230}. Gist understanding is also related to the general problem of image classification \cite{russakovsky2014imagenet}: significant progress has been made in recent years on this task by combining neural approaches like Deep Convolutional Neural Networks with large datasets of labeled images \cite{Krizhevsky:2017} and background knowledge \cite{Deng14}. In our work, we follow this broad line of research, namely linking images to a sense-aware concept repository organised in a taxonomic way (e.g., ImageNet or Wikipedia categories), but do not make any assumption on the vocabulary of the reference knowledge repository like, for instance, the availability of images to be used to train statistical models. This is because, in our case, we want to be able to detect the message of images regardless of the actual topic granularity, namely for concrete as well as abstract (i.e., \emph{non-depictable}) topics. ImageNet, for instance, does not contain images for abstract concepts like `climate change' or `philosophy', for which representative images are hard to collect. To identify gists for concepts which no training data exists, e.g., due to their level of abstractness, we opt instead to learn models on the basis of features extracted from \emph{image-caption-specific knowledge graphs}, as opposed to standard supervised learning from image examples.

Our work crucially builds upon previous contributions from many other fields ranging from multimodal content analysis all the way through entity-based information access and knowledge-based text understanding. Specifically, our work touches on different research topics that stem from the problems of object detection from images, entity linking and retrieval, and using the structure and content of knowledge bases to quantify semantic relatedness and detect topics. We begin our overview of related work with a brief review of work on \emph{multimodal content understanding}, since we are working with data that include both images and short texts, i.e., captions. One modality of the data are images: to lower the effort of manually annotating images with object labels, we rely on \emph{object detection} as pre-processing step, a topic for which we also present a brief review of research contributions most related to our work. Complementary to image processing, our pipeline connects entity mentions, i.e., words and phrases within captions, with concepts from a knowledge base, a problem addressed by the task of \emph{entity linking}. We cast gist detection as an \emph{entity retrieval} problem with an image-caption pair as the query, and accordingly touch upon previous efforts from the IR community for this task. In order to find additional relevant concepts from the knowledge base, we rely on measures of \emph{entity relatedness}. 
Entity relatedness measures have originally been developed for the related, yet different task of \emph{topic and document cluster labeling}.


\bigskip \noindent \textbf{Multimodal content understanding.}
Recent years have seen a growing interest for interdisciplinary work, which aims at bringing together vision and language. 
As such processing of visual data, i.e., video or images, are combined with NLP and text mining techniques. 
This is perhaps no surprise, since text and vision are expected to provide complementary sources of information, and their combination is expected to produce better, grounded models of natural language meaning \cite{bruni12}, as well as enabling high-performing end-user applications \cite{NikolaosAletras2012}.

However, while most of the research efforts so far concentrated on the problem of image-to-text and video-to-text generation -- namely, the automatic generation of natural language descriptions of images \cite{feng10a,kulkarni11,yang11,gupta12}, and videos \cite{barbu12,das13a,krishnamoorthy13} -- few researchers focused on the complementary, yet more challenging, task of associating images or videos to arbitrary texts -- \cite{feng10b} and \cite{das13b} being exceptions. However, even these latter contributions address the arguably easier task of generating literal descriptions of depictable objects found within standard news text, thus disregarding other commonly used, yet extremely challenging, dimensions of image usage such as media icons~\cite{Perlmutter2004,Drechsel2010}. 
Most of non-literal image-text usages has not received much attention in the field of automatic language and vision processing: researchers in Natural Language Processing, only recently started to look at the problem of automatically detecting metaphors \cite{shutova13}, whereas research in computer vision and multimedia processing did not tackle, to the best of our knowledge, the problem of iconic images at all.

\bigskip \noindent \textbf{Object detection from images.}
Supported by the availability of benchmark collections for image retrieval \cite{ThomeeP12} and benchmarking tasks \cite{russakovsky2014imagenet}, a large body of works has focused in the past years on the problem of detecting objects in images \cite[\emph{inter alia}]{Everingham2010,502,Ordonez2011}. These either train object detectors from images with bounding box annotations, learn sparse representation of images~\cite{Shivani2002}, use semantic segmentation, namely an assignment of class labels to pixels~\cite{Du2017c}, or rely on captions to guide the training or generate captions for images, based on an unsupervised model from the spatial relationship of such bounding boxes \cite{Elliott15}. Alternatively, images that are already annotated can be used to find similar images. For this, deep learning approaches \cite{Novak2015} or approaches based on autoencoders \cite{Du2017} have been shown to be achieve state-of-the art performance.

Since many images are accompanied by captions, approaches have been devised that use text in such captions to aid the detection of objects and actions depicted in the image. This idea is exploited using supervised ranking \cite{hodosh2013imagedescription}, using entity linking and WordNet distances \cite{Weegar14}, and using deep neural networks \cite{socher2014grounded}. 
One application is image question answering \cite{Ren2015imageqa}. 
Research to this end has thus far focused on literal image-caption pairs, where the caption enumerates the objects visible in the image.  In contrast, the emphasis of this work is on non-literal image-caption-pairs with media-iconic messages, which allude to an abstract gist concept that is not directly visible. 

Crucial for training object detectors are training data, which ideally consists of images with bounding boxes and affiliated textual labels. 
Image or multimodal datasets either provide a limited vocabulary or only a small fraction of bounding boxes. 
As such ImageNet, even though it provides over 14 million images, has bounding boxes only for 8\% of its images. 
The lack of such training material is the only barrier for application in our domain. For this reason and to facilitate reproducibility of our research, we simulate object detection or rely on an external system such as the Microsoft API. While this work builds on object detection tags, it has been shown that object classes available in ImageNet are insufficient to capture objects found on images on topics of global warming  \cite{weiland2015imagemessage}.

\bigskip \noindent \textbf{Entity linking.}
Detecting entity mentions in text and linking them to nodes in a knowledge base is a task well studied in the TAC KBP venue. 
Most approaches include two stages. The first stage identifies candidate mentions of entities in the text with a dictionary of names. 
These candidates are next disambiguated using structural features from the knowledge graph, such as entity relatedness measures \cite{ceccarelli2013relatedness_entitylinking,Hulpus2015pathbased} and other graph walk features \cite{talukdar2008weaklyrelationkb}.
A prominent entity linking tool is the TagMe! system \cite{ferragina2010tagme}. 
A simpler approach, taken by DBpedia spotlight~\cite{Mendes:2011}, focuses on unambiguous entities and breaks ties by popularity. We evaluate both approaches in Section~\ref{experiments}. 

\bigskip
\noindent \textbf{Entity retrieval.}
Entity retrieval tasks have been studied widely in the IR community in INEX and TREC venues \cite{demartini2009inex,balog2010entitytrec}. The most common approach is to represent entities through textual and structural information in a combination of text-based retrieval models and graph measures \cite{zhiltsov2015fielded}.

Different definitions of entities have been explored. Recently, the definition of an entity as ``anything that has an entry on Wikipedia'' has become increasingly popular. Using entities from a knowledge base that are (latently) relevant for a query for ad hoc document retrieval has lead to performance improvements 
\cite{dalton2014eqfe,kurland2016docentities}.
Moreover, using text together with graphs from article links and category membership for entity ranking has been demonstrated to be effective on freetext entity queries such as "ferris and observation wheels"
~\cite{Demartini2008semantically}.
In contrast to this previous work, our paper focuses on a graph expansion and clustering approach.

%
In order to facilitate robust ranking behaviour, clustering is often combined into a back-off or smoothing framework. 
This has been successfully applied for document ranking by Raiber et al.~\cite{raiber2013clustermrf}, and our approach adopts it for the case of entity ranking.


\bigskip
\noindent \textbf{Entity relatedness.}
The purpose of entity relatedness is to score the strength of the semantic association between pairs of concepts or entities. 
The research on this topic dates back several decades~\cite{ZhangGC13}, and a multitude of approaches have been researched. 
Among them, we place particular emphasis on measures that use a knowledge base for computing relatedness. 
We distinguish two main directions: (i) works that use the textual content of the knowledge base~\cite{Gabrilovich2007computing,Hoffart2012kore}, particularly Wikipedia, and (ii) works that exploit the graph structure behind the knowledge base, particularly Wikipedia or Freebase hyperlinks~\cite{Milne2008learning}, DBpedia~\cite{Schuhmacher2014knowledge,Hulpus2015pathbased}. 
Hulpu{\c{s}} et al.\  \cite{Hulpus2015pathbased} introduced an exclusivity-based measure and found that it works particularly well on knowledge graphs of categories and article membership (which we use also) for modeling concept relatedness. It was shown to outperform simpler measures that only consider the length of the shortest path, or the length of the top-k shortest paths, as well as measures that take into account the semantic relations found along these paths~\cite{Schuhmacher2014knowledge}.

\bigskip \noindent \textbf{Topic and document cluster labeling.}
Other research directions that are closely related to ours are concerned with labeling precomputed topic models~\cite{Mei2007automatic,hulpus2013} and with labeling document clusters~\cite{Carmel2009enhancing}. 
Topic model labeling is the task of finding the gist of a topic resulted from probabilistic topic modeling. 
Solutions to these related problems make implicit or explicit use of knowledge about words and concepts collected from a document corpus. 
Such knowledge is not available for our problem, consequently, most of these approaches are inapplicable for understanding the gist. 

\bigskip \noindent
The work that is most similar to ours is presented in \cite{hodosh2013imagedescription}, who also view image understanding as a ranking task. However, their approach focuses on automatic image description using natural language sentences, as opposed to entity ranking in a knowledge base. Moreover, their goal is to produce \emph{concrete conceptual descriptions} of images, as opposed to our (arguably, more general) problem of gist detection that corresponds, following \cite{dvmmPub58}'s terminology, to detecting \emph{abstract scenes}, i.e., what an image as a whole represents.

%

\section{The Problem of Image Gist Understanding }
\label{problemStatement}

\begin{figure*}[tb]
\centering
\begin{subfigure}{.33\textwidth}
  \centering
  \includegraphics[width=.8\linewidth]{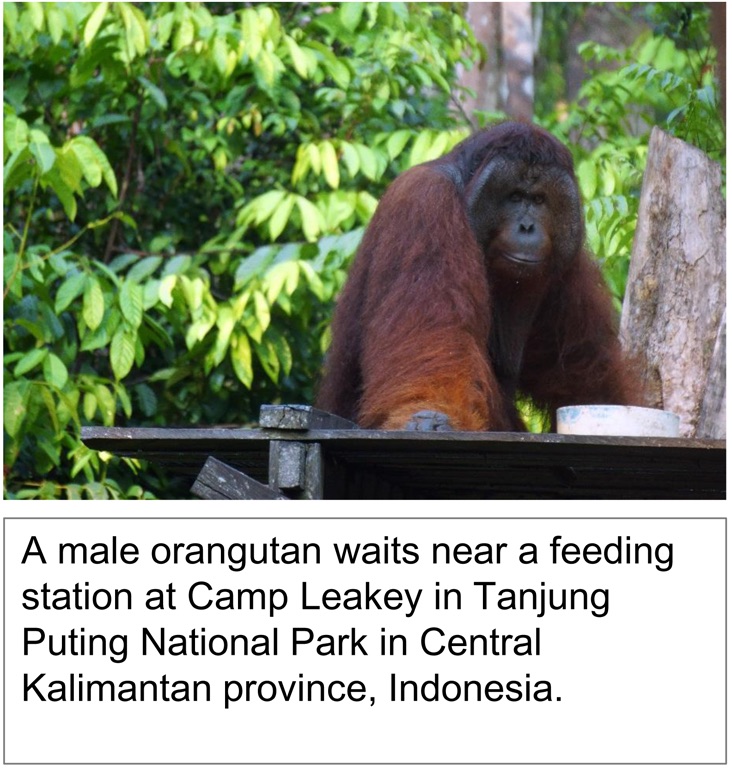}
  \begin{scriptsize}
  \gist{Bornean Orangutans}, \gist{Trees},
  \gist{Mammals of Southeast Asia}, \gist{Botany}, \gist{Plants}
\end{scriptsize}
\caption{Literal Pairing}
  \label{fig:literalexample}
\end{subfigure}%
\begin{subfigure}{.33\textwidth}
  \centering
  \includegraphics[width=.8\linewidth]{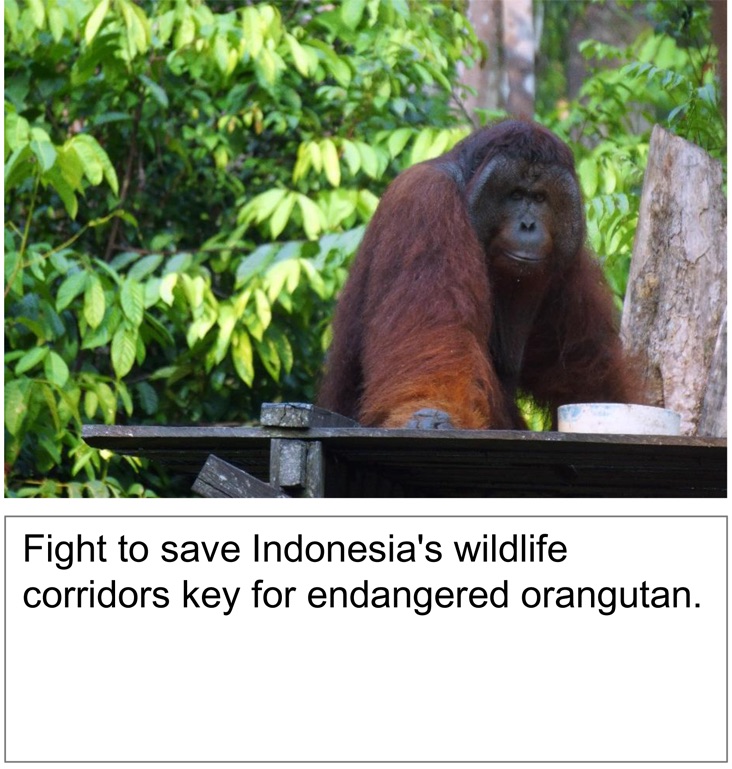}\\
  \begin{scriptsize}
  \gist{Habitat Conservation}, \gist{Biodiversity},
\gist{Extinction}, \gist{EDGE species}, \gist{Deforestation}
  \end{scriptsize}
  \caption{Non-literal Pairing}
  \label{fig:nonliteralexample}
\end{subfigure}
\begin{subfigure}{.33\textwidth}
  \centering
  \includegraphics[width=.8\linewidth]{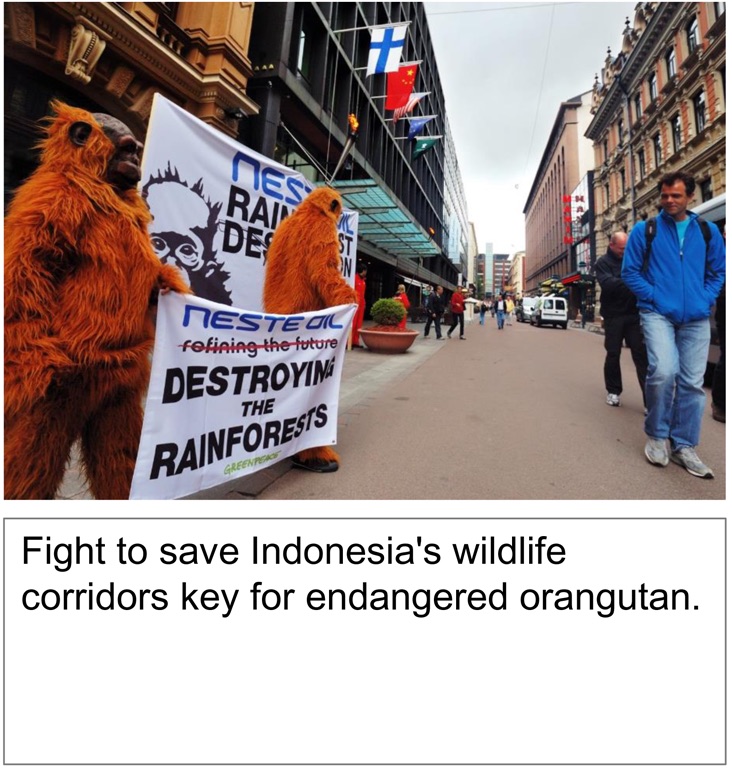}\\
      \begin{scriptsize}
\gist{Habitat Conservation}, \gist{Biodiversity},
 \gist{Extinction}, \gist{Politics},
\gist{Protest}
  \end{scriptsize}
  \caption{Non-literal Pairing}
  \label{fig:nonliteralexample2}
\end{subfigure}
\caption{Example image-caption pairs sharing either images or captions with their respective gist nodes
\scriptsize{}(a, b, c: http://reut.rs/2cca9s7, REUTERS/Darren Whiteside, a, b: http://bit.ly/2nGIa58, Flickr/Budi Nusyirwan, CC BY-SA 2.0, c: http://bit.ly/2p3Y7n4, Commons/Lauri Myllyvirta, CC BY 2.0, last accessed: 04/06/2017.)}
\label{fig:example}
\end{figure*}

\noindent
Our goal is to understand the gist conveyed by a given image-caption pair. In this work we make a first step in this direction by algorithmically identifying which concepts in a knowledge base describe the gist best. We cast the task of gist detection as a \emph{concept ranking problem} -- namely, to predict a ranking of concepts (i.e., Wikipedia articles and categories) from a knowledge base ordered by their suitability to express the gist of a given image-caption pair.

\bigskip\noindent\textbf{Task Definition:} Predict a ranking of concepts from the knowledge base ordered by their relevance to express the message (gist) of an image-caption pair.

\bigskip\noindent\emph{Given:} An image and its associated textual caption, for which the gist needs to be found. Furthermore, given a knowledge base, which is viewed as a graph consisting of a vocabulary of concepts (i.e., nodes) and semantic relations between them (i.e., edges). Additionally, given the textual descriptions associated to concepts and entities.

\bigskip\noindent\emph{Output:} A ranked list of concepts expressing the message conveyed by the image.

\bigskip\noindent\textbf{Terminology: seed vs.\ gist knowledge base nodes.} In order to leverage the content and structure of the knowledge base, a link between the objects that are visible in the image, the linguistic expressions (i.e., nouns, proper names) found in the caption and their corresponding concepts in the knowledge base is established using so-called \emph{linking} methods: we call the corresponding nodes in knowledge graph \emph{seed nodes}. Here, we aim to rank the nodes of the knowledge graph based on their relevance to the seed nodes, thus, the initial query. The highly ranked ones then become the gist of the image-caption pair.  A node that corresponds to the gist of an image-caption pair is referred to as \emph{gist node}. As a consequence, in this work a gist concept can refer to any node in a given knowledge graph, envisioning any of the general-purpose knowledge bases that are congruent to Wikipedia (including DBpedia \cite{Bizer09}, YAGO \cite{hoffart13}, etc.).

\bigskip\noindent\textbf{Beyond literal meaning: literal vs.\ non-literal image-captions.} We define an image and its affiliated textual caption as \emph{image-caption pair.}
We define further two types of pairs, depending on the kind of message they convey.
\emph{Literal} pairs are those in which the caption describes or enumerates the objects depicted in the image.
\emph{Non-literal} pairs, of which media icons are an example, are those which typically convey an abstract message and where images and captions often contain complementary information.

We claim that in order to understand the gist of images, both the image and the caption are needed.
As they together form a union, an image-caption pair can encode a different gist by changing the caption.
Vice versa, combining a caption with a different image can shift the focus of the gist or change the semantics.
To illustrate the effects of different image-caption pairs, we show three different pairs as examples in Figure~\ref{fig:example}.
Two of the pairs are media icons commonly used to convey the message of species threatened by deforestation. One is a literal pair (cf. Figure~\ref{fig:literalexample}), which lacks the connection to threat, extinction, and deforestation.
The caption describes the image showing an orangutan in what seems to be a national park. The gist of the pair is
\gist{Bornean Orangutan}.


By exchanging the caption it becomes apparent that the gist is habitat conservation to save an endangered animal (cf. Figure~\ref{fig:nonliteralexample}).
Considering the corresponding caption thus helps in the disambiguation of the gist.
On the other hand, captions alone are often brief, and when taken out from the context of the image, they fail to convey the entire gist.
For instance, by inspecting only the caption \word{Fight to save Indonesia's wildlife corridors key for endangered orangutan}, it is not clear whether the focus is on endangered species as victims of deforestation, as depicted Figure~\ref{fig:nonliteralexample}, or on people who fight for habitat conservation, as depicted in Figure~\ref{fig:nonliteralexample2}. That is, only an image can disambiguate the gist.
We consequently consider an image-caption pair as the targeted \emph{query} for which gist concepts are ranked.
\section{Methodology}\label{approach}

\noindent The main idea behind our approach is to use a general-purpose knowledge base to understand the message conveyed by an image and its caption. To this end, we develop a framework for gist detection based on the following pipeline: First, detected objects in the image and entity mentions in the caption are linked to a reference machine-readable repository of knowledge. Our hunch is to exploit the content and connectivity of the knowledge base, which we view as a graph (hence a `knowledge graph' \cite{Paulheim17}), in order to identify relevant topics that capture not only the content of the image-caption pair, but also its intended meaning. By using the knowledge base as a graph, we can represent the concepts collected through object detection (in the image) and entity detection (in the caption) as nodes. Next, the neighborhood of these projected nodes in the knowledge graph is inspected to provide a set of candidate gists. Finally, we combine (1) content-based features, extracted from the textual descriptions of the knowledge base entities, and (2) graph-based features obtained by analyzing the structure of the knowledge graph. Lastly, these features are combined into a node ranking model that pinpoints the gist concepts for a given image-caption pair.

The key novel aspects of our approach are:

\begin{itemize}[leftmargin=4mm]

\item We present a complete approach to perform \emph{knowledge-based image gist detection}. At the heart of our method lies the idea that we can leverage the content and structure of a knowledge base to identify concepts to understand the, possibly abstract, message conveyed by an image-caption pair as a whole.

\item We propose a methodology that views \emph{knowledge-aware image classification as entity ranking}. Our hunch is that, given a knowledge graph that covers the subject of the image-caption pair, the gist concepts lie in the proximity of those mentioned in the caption or depicted in the image, namely the seed nodes. We define features of candidate gist nodes based on their graph relations or textual content according to their corresponding concept page in the knowledge base. These, in turn, are used to build a supervised ranking model that is able to rank concepts on the basis of their relevance for the image-caption pair.

\item By using background knowledge from the knowledge base, \emph{our method is able to identify gist concepts that are neither visible in the image, nor necessarily explicitly mentioned in the caption}. We expect this hypothesis to be true especially for pairs with an abstract gist, e.g., media icons.  Examples of such concepts transmitting the message referred to as gist are \gist{global warming}, \gist{endangered species}, \gist{biodiversity}, or \gist{sustainable energy}\footnote{%
  We use \textsf{Sans Serif}\ for words and queries,
  \textsc{Small Caps} for gists, Wikipedia pages and categories.%
}.
Despite not being depictable and consequently identifiable by image recognition, the gist nodes will likely be in close proximity in the knowledge graph to the objects in the image that are visible, as well as to the concepts mentioned in the captions.

\end{itemize}


\begin{figure*}[t]
\begin{center}
 \includegraphics[width=\textwidth, height=120.0pt]{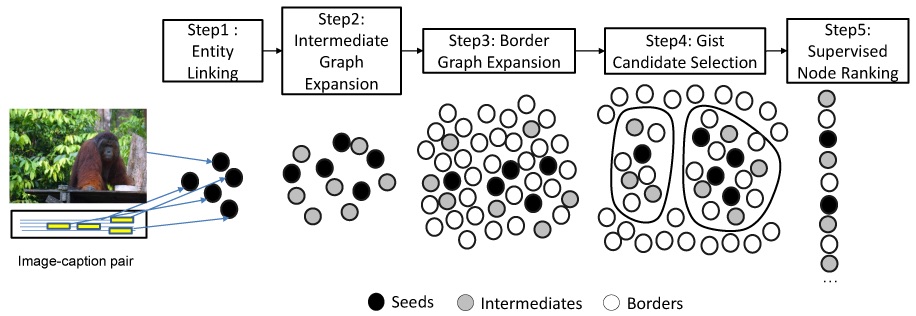}
 \caption{Our gist extraction and ranking pipeline (edges between nodes removed for simplicity).}
  \label{fig:pipeline}
\end{center}
\end{figure*}

We present our approach as a pipeline (Figure~\ref{fig:pipeline}). For explanatory purposes, we make use of the media icon of Figure~\ref{fig:nonliteralexample} to provide us with a running example to illustrate each step of our pipeline.

\subsection{The Knowledge Graph}\label{subsect:kg}
\noindent As a preliminary for our pipeline, we need to represent the knowledge base as knowledge graph, to benefit from semantic information like, for instance, knowledge-based semantic relatedness between two entities or concepts, that have been shown to perform well on tasks such as text understanding \cite{ZhangGC13}.

Given a knowledge base, we define a \emph{knowledge graph} \cite{Paulheim17} as the directed or undirected graph $\mathbf{KG}(V,E,T)$ such that the set of nodes $V$ contains all concepts in the knowledge base, every edge $e_{ij} \in E$, $E \subseteq V \times T \times V$ corresponds to a relation in the knowledge base between two nodes $v_i$ and $v_j$, and the set $T$ defines the relation types in the knowledge base. For our purposes, we consider the knowledge graph to be undirected, unless specified otherwise. Additionally, we denote labeled edges in the graph as $(v_i,t,v_j)$, which is assumed to imply: i) $v_i,v_j \in V$, ii) $(v_i,t,v_j) \in E$, iii) $t \in T$.

In this work, we opt for Wikipedia, since it provides a wide-coverage, general-purpose knowledge base \cite{hovy13a} containing large amounts of manually-curated text describing millions of different entities across a wide range of heterogeneous domains. Furthermore, and perhaps even more importantly for our approach, the link structure of Wikipedia can be exploited to identify topically associative nodes, thus making it possible to complement the textual content of the knowledge base with that derived from the structure of the underlying knowledge graph. However, our method can be also used with any other lexical or ontological resource, e.g.\ YAGO \cite{hoffart13} or DBpedia \cite{Bizer09}, provided it can be cast as a knowledge graph containing disambiguated entities with explicit semantic relations and textual descriptions.

Our knowledge graph contains as nodes all articles and categories from the English Wikipedia, similarly, e.g., to \cite{FlatiVPN16}. As for edges, we consider the following types of relations $T$, which have been previously found to provide useful information for topic labeling \cite{hulpus2013}:
\begin{itemize}[leftmargin=4mm]


\item \textbf{Page-category links}: The category membership relations that link an article to the categories it belongs to (e.g., page \textsf{Wildlife corridor} is categorized under \textsc{Wildlife conservation}). These relations provide topics for different aspects of the concepts described by the Wikipedia page.

\item \textbf{Super- and sub-category links} The relationship between a category and its parent category (e.g., \textsc{Wildlife conservation} is a sub-category of  \textsc{Conservation}), as well as its children categories (e.g., \textsc{Conservation} is a super-category of \textsc{EDGE species}). Relations between categories can be taken to capture a wide range of topical associations between concepts of different granularities \cite{nastase12}, including semantic generalizations and specializations \cite{ponzetto11a}.

\end{itemize}





\subsection{Step 1: Image and Caption Node Linking}
\noindent Initially, we project objects depicted in the image and concepts mentioned in the caption onto nodes in the knowledge base. That is, given an image-caption pair $(C,I)$, we want to collect a set of \emph{seed nodes} $S$ in the knowledge graph (Section \ref{problemStatement}). Here, we view the caption as a set of \emph{textual mentions} $C = \{m_1, \ldots, m_n\}$, namely noun phrases that can be automatically extracted using a standard NLP pipeline (in this work, we use the StanfordNLP toolkit~\cite{Manning14}). Next, each mention $m \in C$ is linked to a corresponding concept from the knowledge graph. The linking can be achieved with different strategies (cf.\ Section~\ref{rq1} for the description and evaluation of different linking strategies).

The linking of $m$ to a concept is defined as a node $v_c$ in the knowledge base such that $v_c \in V_{C} \, \cup \, \epsilon \subset V$, namely the union between the set of \textbf{caption nodes} $V_{C}$ and an `undefined concept' $\epsilon$ if no linking is possible -- i.e., there is no corresponding concept for $m$ in the knowledge base. The image is viewed as consisting of a set of \emph{object labels} $I = \{l_1, \ldots, l_n\}$ that are either manually given, or are taken from the output of an automatic object detector. Similarly to the captions' mentions, each of the labels  $l \in I$ needs to be linked to a corresponding concept $v_i$ from the knowledge graph such that $v_i \in V_{I} \, \cup \, \epsilon \subset V$, namely the union between the set of \textbf{image nodes} $V_{I}$ and, again, the `undefined concept' $\epsilon$ when no concept in the knowledge base is available corresponding to the meaning of $l$. Finally, we take the union of mapped textual mentions and object labels, namely caption and image nodes as the set of \textbf{seed nodes} -- i.e., the concepts from the knowledge base that corresponds to the entities and objects found in the image-caption pair:

\begin{equation}
S = V_C \cup V_I, \quad S \subset V
\end{equation}

\noindent There exist many different ways to link string sequences such as textual mentions (from the caption) and  object labels (from the images) to concepts in a knowledge base -- i.e., the so-called problem of \emph{entity linking}, which has received much attention in recent years (Section \ref{rw}). Here, we opt for a simple iterative concept linking strategy that is both applicable to captions' mentions and images' object labels, and is particularly suited for short object labels for which no textual context is available to drive the disambiguation process. First, we attempt to link mentions and labels to those Wikipedia articles whose title lexicographically matches, e.g., \wikipage{Indonesia}. Additionally, whenever we find a title of a disambiguation page \wikipage{Orangutan (disambiguation)}, we include all redirected articles that can be reached with two hops at most from previously linked nodes along the Wikipedia graph. In our experiments (Section \ref{experiments}), we demonstrate that this simple approach is, for the purpose of our task, as good as TagMe \cite{ferragina2010tagme}, a state-of-the-art entity linking system.

\bigskip \noindent \textbf{Example.} In our working example (Figure~\ref{fig:nonliteralexample}), objects in the image have been associated with labels like \word{orangutan}, \word{sign}, \word{trunk}, \word{tree}, \word{ground}, and \word{vegetation}. The NLP pipeline, instead, extracted mentions from the caption like  \word{fight}, \word{Indonesia}, \word{wildlife}, \word{corridor}, \word{key}, and \word{orangutan}. These, in turn, are linked to seed nodes such as \wikipage{Indonesia} and \wikipage{Wildlife corridor}, among others (Figure~\ref{fig:runningexampleintermed}, depicted in grey).






\subsection{Step 2: Intermediate Graph Expansion}

\begin{figure*}[t]
\centering
\begin{minipage}{0.95\textwidth}
\centering
\begin{subfigure}[t]{0.90\textwidth}
	\includegraphics[width=.9\textwidth]{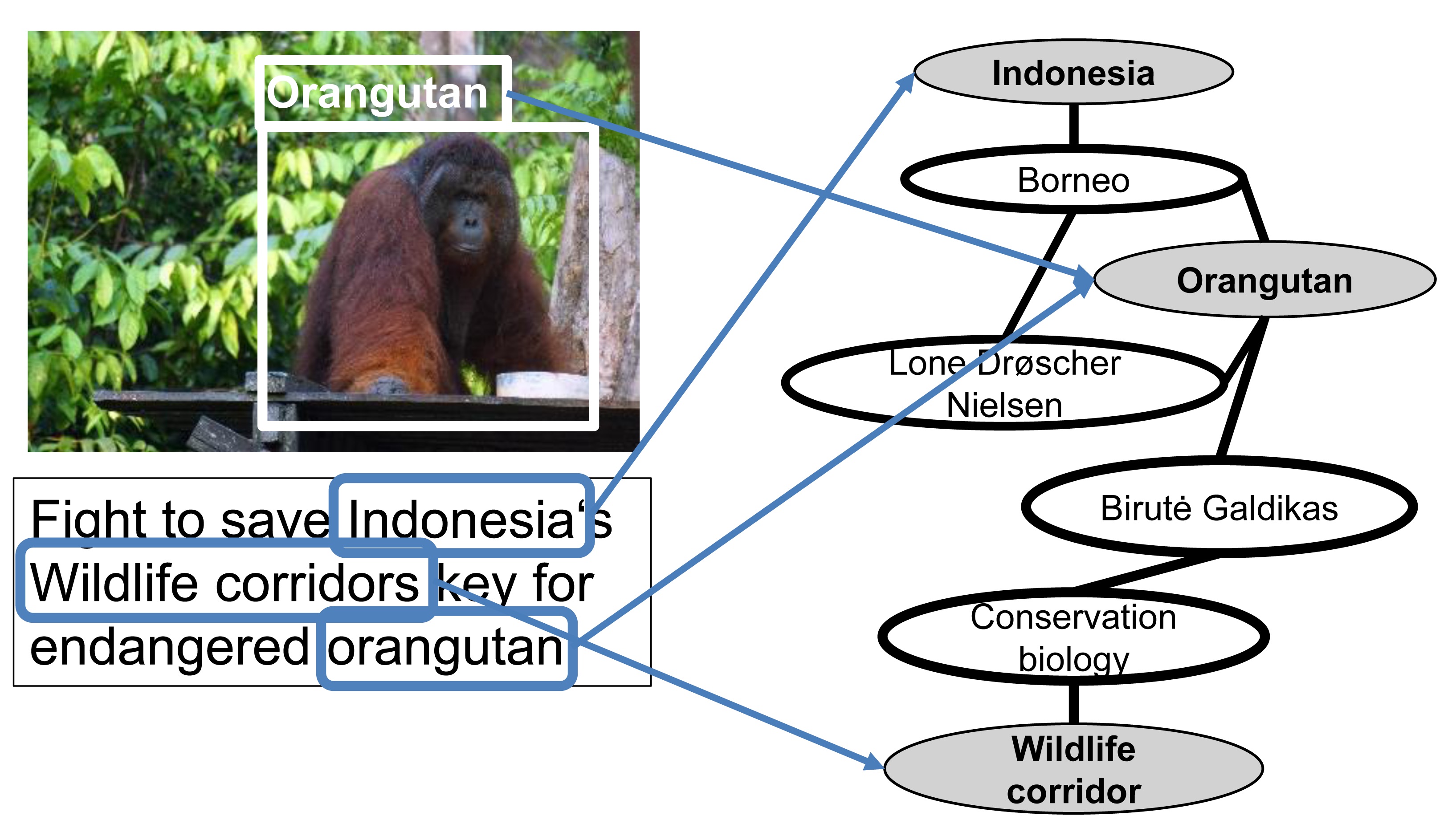}
\end{subfigure}
\\
\begin{subfigure}[t]{0.95\textwidth}\centering
	\includegraphics[width=.98\textwidth]{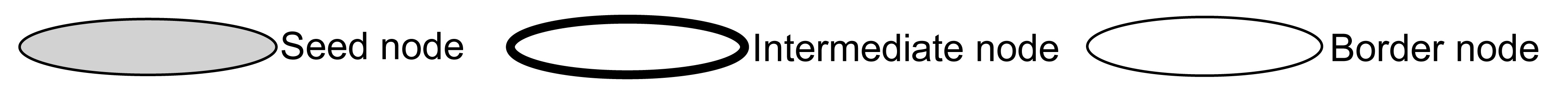}\\
\end{subfigure}
\caption[Intermediate graph example]{Example of intermediate graph for the image-caption pair in Figure~\ref{fig:nonliteralexample}.}
\label{fig:runningexampleintermed}
\end{minipage}
\end{figure*}

\noindent Especially for media-iconic pairs, one cannot assume that the gist corresponds to any of the concepts found among those obtained by linking either the image labels or the textual captions. For instance, in the case of our example (Figure~\ref{fig:nonliteralexample}), we cannot find the gist node \wikicategory{EDGE Species} among any of the seed nodes identified in Step 1 (i.e., those highlighted in grey in Figure~\ref{fig:runningexampleintermed}). That is, Step 1 may not be sufficient to identify such gists by simple entity linking, especially in the case of abstract, non-depictable concepts that are rarely mentioned explicitly in the caption.

We operationalize our hypothesis that gist nodes will be found in the knowledge base on paths between seed nodes as follows. We start with the seed nodes from Step 1 and build a query-specific knowledge graph by extracting all the paths that connect pairs of seeds -- similar in spirit to previous approaches for knowledge-rich lexical \cite{NavigliLapata:10} and document \cite{Schuhmacher2014knowledge} understanding.
More specifically, we start with the seed nodes $S$ and create a labeled directed graph $G_I = (V_I, E_I)$ as follows: a) first, we define the set of nodes $V_I$ of $G_I$ to be made up of all seed concepts, that is, we set $V_I = S$; b) next, we connect the nodes in $V_I$ based on the paths found between them in Wikipedia. Nodes in $V_I$ are expanded into a graph by performing a depth-first search (DFS) along the Wikipedia knowledge graph (Section \ref{subsect:kg}) and by successively adding all simple directed paths $v, v_1, \ldots, v_k, v'$ ($\{v, v'\} \in S$) of maximal length $L$ that connect them to $G_I$, i.e., $V_I = V_I \cup \{ v_1, \ldots, v_k \}$, $E_I = E_I \cup \{ (v,t_1,v_1), \ldots, (v_k,t_k,v') \}$, $t_i \in T$. As a result, we obtain a subgraph of Wikipedia containing the initial concepts (seed nodes), together with all edges and intermediate concepts found along all paths of maximal length $L$ that connect them. In this work, we set $L=4$ (i.e., all paths with length shorter than 4) based on a large body of evidence from previous related work \cite[\emph{inter alia}]{navigli12,hulpus2013,Schuhmacher2014knowledge}. We call the nodes along these paths, except the seed nodes, \textbf{intermediate nodes}, $I = V_I \setminus S$. The graph resulted from combining all the nodes on these paths (including the seeds) as well as the edges of the paths, is what we call the \textbf{intermediate graph}:
\begin{equation}
\mathbf{KG}_I(V_I, E_I, T), \quad V_I = S \cup I
\end{equation}

\bigskip \noindent \textbf{Example.} The graph shown in Figure~\ref{fig:runningexampleintermed} is obtained by connecting three concepts, namely \wikipage{Orangutan}, \wikipage{Indonesia} and \wikipage{Wildlife corridor} with connecting paths found in Wikipedia.

\subsection{Step 3: Border Graph Expansion}
\label{bordergraph}

\begin{figure*}[t]
\centering
\begin{minipage}{0.95\textwidth}
\begin{subfigure}[t]{0.98\textwidth}\centering
	\includegraphics[width=.9\textwidth]{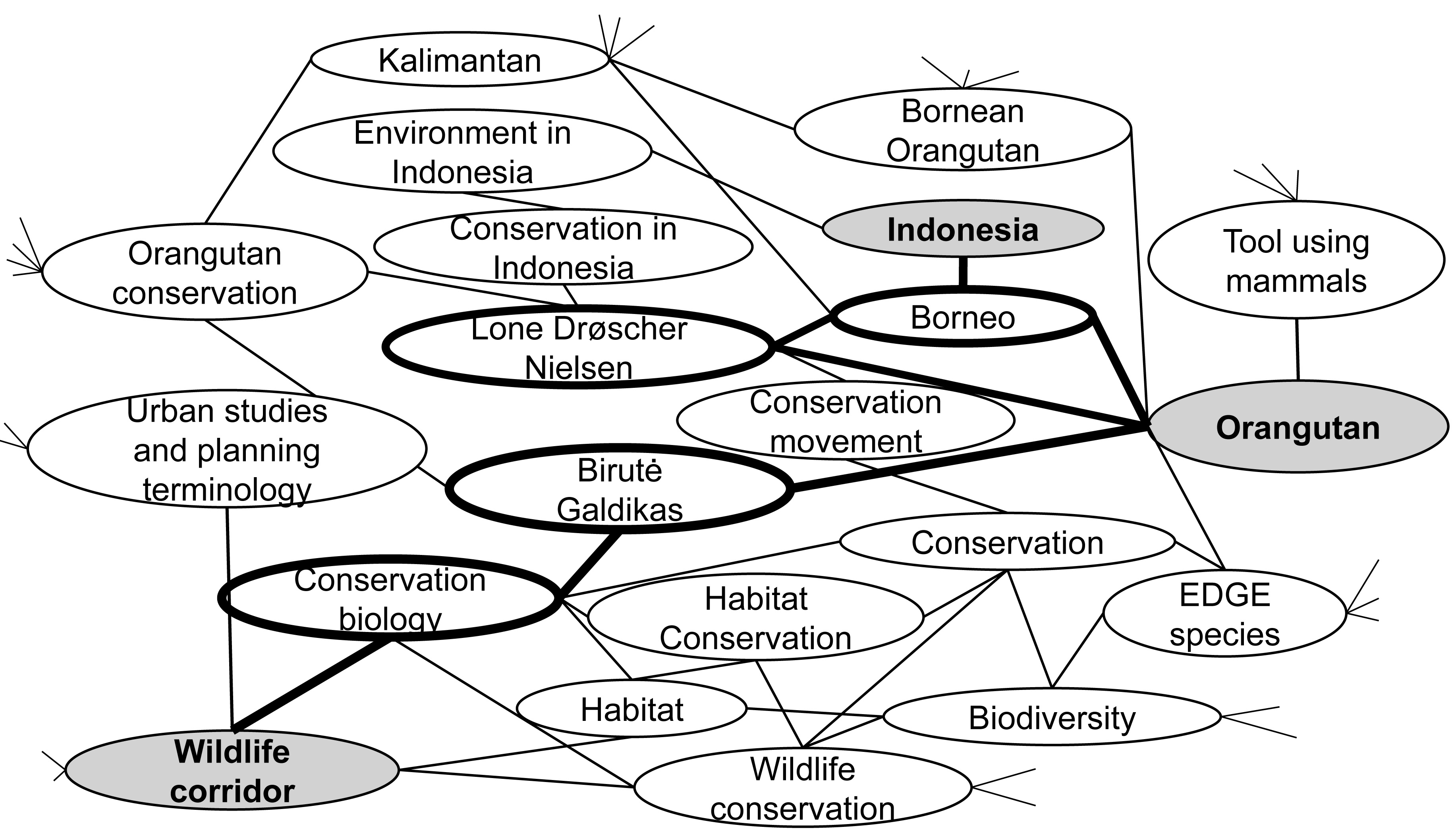}
\end{subfigure}
\\
\begin{subfigure}[t]{0.95\textwidth}\centering
	\includegraphics[width=.98\textwidth]{img/legend.jpg}\\
\end{subfigure}
\caption[Border graph examples]{Border graph example for the image-caption pair in Figure~\ref{fig:nonliteralexample}. For simplicity, the image does not make the distinction between article nodes and category nodes, and it also omits edge directions and edge costs.}
\label{fig:runningexample}
\end{minipage}
\end{figure*}

\noindent The intermediate graph we just built by connecting seed concepts can be used to identify the region of the reference knowledge graph (i.e., Wikipedia)  that covers the topics of the image-caption pair. However, while graphs of this kind have been extensively shown to be useful for text lexical understanding \cite[\emph{inter alia}]{NavigliLapata:10,ponzetto10b,navigli12e}, it might still be the case that they do not contain relevant gist nodes -- e.g., in the graph in Figure~\ref{fig:runningexampleintermed} we cannot find any of the gists  \gist{Habitat Conservation}, \gist{Biodiversity}, \gist{Extinction}, \gist{EDGE species} or \gist{Deforestation} from our example (Figure~\ref{fig:nonliteralexample}). We additionally expand the intermediate graph to include all neighbors and their connecting paths that can be reached within two hops from the nodes it contains.

To expand our semantic graphs we use a procedure similar in spirit to the one we used to create intermediate graphs. We start with the nodes from the intermediate graph $V_I$ and create a labeled directed graph $G_B = (V_B, E_B)$ as follows: a) first, we define the set of nodes $V_B$ of $G_B$ to be made up of all nodes from the intermediate graph by setting $V_B = V_I$; b) next, we expand the set of nodes in $V_B$ using a DFS along Wikipedia, such that for all paths $v, v_1, \ldots, v_k, v'$ ($v \in V_I$, $v' \in V$) of maximal length 2 we set  $V_B = V_B \cup \{ v_1, \ldots, v_k, v' \}$ and $E_B = E_B \cup \{ (v,t_1,v_1), \ldots, (v_k,t_k,v') \}$, $t_i \in T$.  The nodes that are added to the graph by the expansion are called \textbf{border nodes}, as they lie between the seeds, intermediates, and the rest of the knowledge graph, i.e., $B = V_B \setminus V_I$. We name the resulting graph the \textbf{border graph}:
\begin{equation}
\mathbf{KG}_B(V_B, E_B, T), \quad V_B = S \cup I \cup B
\end{equation}
Figure~\ref{fig:runningexample} shows a part of the border graph obtained from the intermediate graph of  Figure~\ref{fig:runningexampleintermed}.

\bigskip \noindent \textbf{Example.}
From the sparse information in the intermediate graph (Figure \ref{fig:runningexampleintermed}), it is already clear that \gist{Indonesia} is much closer than \gist{Wildlife Corridor}. Now, the border graph encloses a semantic relatedness, which might prefer structural further away concepts or penalize structural closer concepts, such as the mentioned \gist{Indonesia} and \gist{Orangutan}, just because they are connected over paths that are semantically less relevant. However, in our example, the structural short distance is in line with the semantic relatedness: we find that \gist{Indonesia} and \gist{Orangutan} are much closer than \gist{Wildlife Corridor} and \gist{Orangutan}.

\subsection{Step 4a: Clustering of Seed and Intermediates}

\begin{figure}[t]
\begin{center}
 \includegraphics[width=0.9\textwidth]{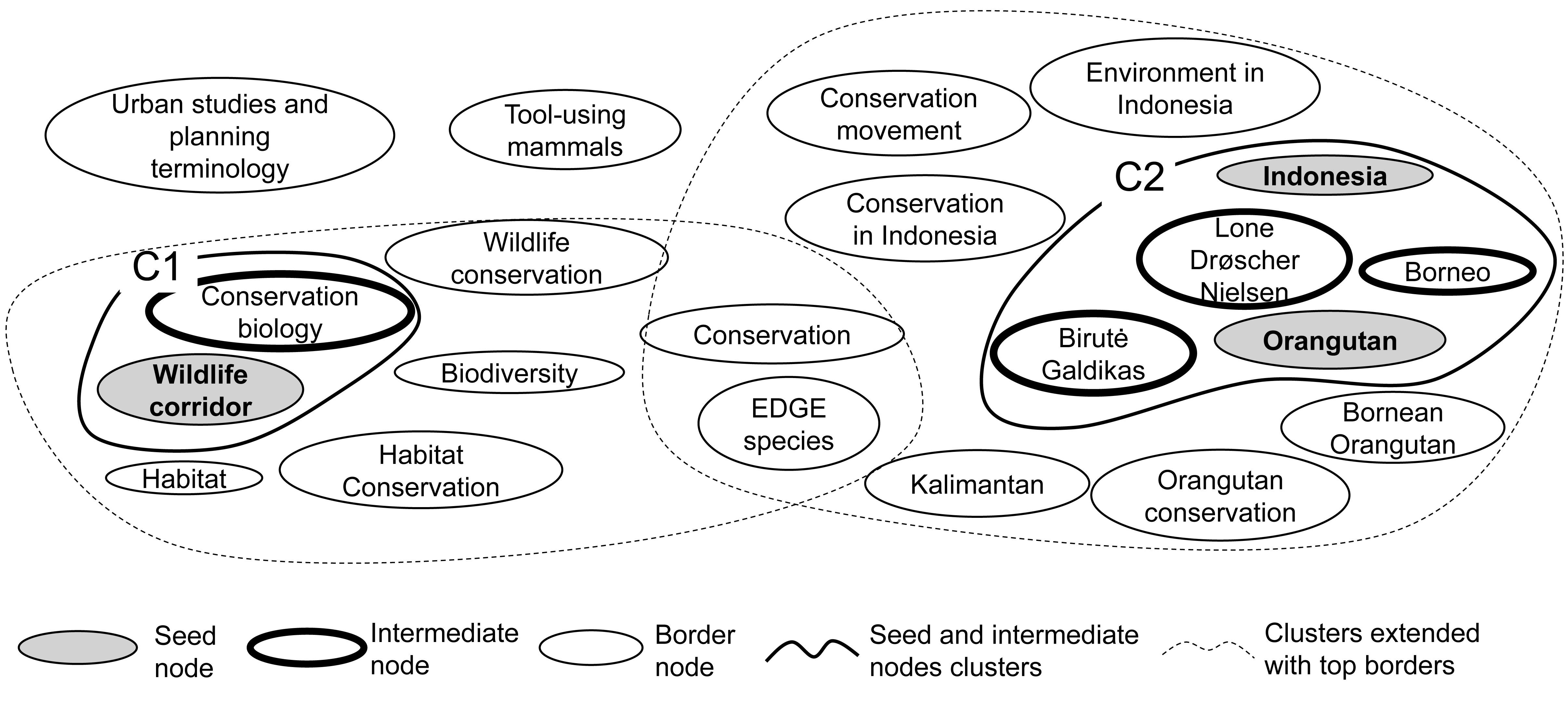}
 \caption{Example of clusters of seeds and intermediates, extended with their most related border nodes (top borders). The border nodes that have only weak semantic associations to the clusters are filtered out (e.g., \wikicategory{Tool-using mammals} and \wikicategory{Urban studies and planning terminology}).}
  \label{fig:extendedclusters}
\end{center}
\end{figure}

\noindent After the previous step, we obtain a graph that contains all the concepts from the image and its caption, as well as other concepts from the knowledge graph that lie in close proximity.
As previously stated, our assumption is that the gist nodes are part of this graph, and that graph properties will make them identifiable. However, a challenge is that often, an image-caption pair covers multiple sub-topics. These sub-topics represent different aspects of the core topic (so-called core gist) of an image-caption pair, e.g., the core gist \gist{Habitat Conservation} has the aspects of habitat conservation in general and region-specific habitat conservation (cf. Fig~\ref{fig:extendedclusters}, expressed by the clusters in dashed lines).
Applying the border graph strategy directly on the seed and intermediate graph in the presence of multiple topics will most often result in a semantic drift and low-quality results.


We identify weakly related sub-topics of an image-caption pair by clustering the set of seed and intermediate nodes - defined as $V_I$ in Step $2$.  To this end, we compute a distance metric $\sigma^{(-1)}: V\times V  \rightarrow \mathbb{R}^+$ for all pairs of nodes in set $V_I$.  Having these pairwise distance scores, we apply the Louvain clustering algorithm~\cite{blondel2008fast}, a nonparametric, modularity optimization algorithm. Any other clustering algorithm can be used that takes as input a pairwise distance matrix.
The resulting clusters group the seeds and intermediates into what we call sub-topics of the image-caption pair.

For completeness, we now briefly describe the used distance metric. Its purpose is to capture the inverse of similarity, relatedness, or semantic association between the concepts that are represented by the nodes in the knowledge graph. A great variety of semantic relatedness measures have been studied~\cite{ZhangGC13}. Here, we follow Hulpu{\c{s}} et al.\ \cite{Hulpus2015pathbased}, who introduce the exclusivity-based measure that we use here as a node metric. The authors found that it works particularly well on knowledge graphs of categories and article membership (which we use also) for modeling concept relatedness. It was shown to outperform simpler measures that only consider the length of the shortest path, or the length of the top-k shortest paths, as well as the measure proposed in~\cite{Schuhmacher2014knowledge}.

The exclusivity-based measure assigns a \emph{cost} for any edge $s\stackrel{r}{\rightarrow}t$ of type $r$ between source node $s$ and target node $t$.
The cost function is the sum between the number of alternative edges of type $r$ starting from $s$ and the number of alternative edges of type $r$ ending in $t$:

\begin{equation}
\text{cost}(s\stackrel{r}{\rightarrow}t) = |\{s\stackrel{r}{\rightarrow}*\}| + |\{*\stackrel{r}{\rightarrow}t\}|-1,
\label{formula:cost}
\end{equation}

\noindent where $1$ is subtracted to count $s\stackrel{r}{\rightarrow}t$ only once.

The more neighbors connected through the type of a particular edge, the less informative that edge is, and consequently the less evidence it bears towards the relatedness of its adjacent concepts.
By summing up the costs of all edges of a path $p$, one can compute the cost of that path, denoted $\text{cost}(p)$.
The higher the cost of a path, the lower its support for relatedness between the nodes at its ends.
Thus, given two nodes, $s$ and $t$, their relatedness is computed as the inverse of the weighted sum of the costs of the top-$k$ shortest paths between them (ties are broken by cost function).
Each path's contribution to the sum is weighted with a length based discounting factor $\alpha$:
\begin{equation}
\sigma(s,t)=\sum_{i=1}^k\alpha^{\text{length}(sp_i)}\times \frac{1}{\text{cost}(sp_i)}
\label{formula:relatedness}
\end{equation}
\noindent where $sp_i$ denotes the $i$'th shortest path between $s$ and $t$,
$\alpha \in (0,1]$ is the length decay parameter and $k$ is a number of shortest paths to consider. See \cite[Figure 1]{HulpusPH15} for an example of how this exclusivity-based measure is computed.

In this work, we apply the distance metric resulted as the inverse of the exclusivity based relatedness measure to all pairs of seeds and intermediates, over the border graph.

For clustering the seeds and intermediate nodes, as mentioned, we use the Louvain clustering algorithm. It is a modularity optimization algorithm, therefore it will try to maximize the density of edges inside clusters to edges outside clusters. Since in our case the edges are weighted by semantic distance, the algorithm tries to create clusters of seeds and intermediates, such that the overall distance within one clusters is lower than the overall distance between clusters.
As such, clustering results in groups of seed and intermediate nodes $C = \{ C_1, \dots, C_n \}$ ($C_i \subseteq S \cup I$) that broadly correspond to different sub-topics of the image-caption pair. For more details over modularity-optimization clustering, we refer the interested reader to \cite{blondel2008fast}.




\bigskip \noindent \textbf{Example.} Figure \ref{fig:extendedclusters} shows two clusters identified for our example (Figure~\ref{fig:runningexample}): $C_1$ is about wildlife conservation, containing the seed node \wikicategory{Wildlife Corridor} and intermediate \wikicategory{Conservation Biology}, and $C_2$ covers instead topics about Indonesia, including two seed nodes and three intermediates.


\subsection{Step 4b: Selecting Gist Candidates}

\noindent In the next step, we identify suitable border nodes that make good gist candidates. We hypothesize that these are the border nodes that are close to any of the clusters according to the distance metric $\sigma$. We therefore compute for every border node $x \in B$ its average distance $\bar{\sigma}$ to each cluster $C_i \in C$:
\begin{equation}
\bar{\sigma}(x,C_i) = \frac{1}{\left\vert{C_i}\right\vert}\sum_{y\in C_i}{\sigma(x, y)}
	\label{formula:score}
\end{equation}
where $\sigma$ is the same distance metric we used in the previous step, namely the inverse of the exclusivity-based semantic relatedness measure of \cite{Hulpus2015pathbased}.

For each cluster $C_i$, we select its candidate border nodes as the top-k scoring concepts $Gist_{C_i}$. The final set of candidate gist nodes is built as the union of the top-k border nodes across all clusters, together with the set of seed and intermediate nodes:
\begin{equation}
 Gist = \bigcup_{C_i \in C} Gist_{C_i} \cup S \cup I
	\label{formula:gist}
\end{equation}
These nodes constitute the candidate node set which is ranked in the following step.


\bigskip \noindent \textbf{Example.} The association of top-border nodes with the two example clusters is illustrated in Figure \ref{fig:extendedclusters}. For instance, the wildlife cluster $C_1$ includes \gist{Habitat} and \gist{Biodiversity}, whereas both \gist{Orangutan Conservation} and the the geographic region \gist{Kalimantan} are associated with cluster $C_2$. The border node \gist{Conservation} is associated with both clusters. These border nodes are included in the candidate set, in contrast to borders with a high distance such as \gist{Urban studies and planning terminology} which are left out.

\subsection{Step 5: Supervised Node Ranking}\label{sec:step5}

\noindent For our task, we train a supervised learning model on labeled data, a method which has been shown to provide robust performance across a wide range of information retrieval and natural language processing tasks  \cite{li2011letor}. Moreover, it provides us with a clean experimental setting to evaluate the contribution of different information sources (i.e., relevance indicators).

The objective of the learning-to-rank method is then to learn a retrieval function such that the computed ranking scores produce the best possible ranking according to some evaluation or loss function. For each of the candidate nodes among those found in the set $Gist$, a feature vector $x$ is created and ranked for relevance with supervised learning-to-rank. Many of our features rely on the topography of the graphs we built as part of our pipeline, including node \emph{degree} and \emph{local clustering coefficient}~\cite{Griffiths07topicsin}, as well as graph centrality measures like PageRank~\cite{BrinPage:98} and betweenness centrality~\cite{Freeman78centralityin}. Consequently, the feature vector consists of the features listed in Table \ref{tab:features} collected from the various steps of the pipeline:

\begin{table}[t]
\caption{Features for supervised re-ranking \label{tab:features} }
\scriptsize
\begin{center}

\begin{tabular}{llcccccc}
\toprule
&Feature  & Pipeline  & \multicolumn{5}{c}{Feature set type} \tabularnewline
& & Step & seed & intermediate & border & other & baseline  \tabularnewline
\midrule
1. & is seed node?  & 1 & \checkmark & & & & \\

2. & is intermediate node?  & 2  & & \checkmark & & & \\

3. & Page Rank on intermediate graph  & 2  & & \checkmark & & & \\

4. & Betweenness centrality on   & 2  & & \checkmark & & & \\
	 &			intermediate graph					&   & & & & & \\
5. & is border node?  & 3  & & & \checkmark & & \\

6. & max node-cluster relatedness  & 4  &  & & \checkmark & & \checkmark \\

7. & avg node-cluster relatedness  & 4  &  & & \checkmark & & \\

8. & sum node-cluster relatedness  & 4  &  & & \checkmark & & \\

9. & is member of cluster with   & 4  &  & & \checkmark& & \\
& most seed nodes?  &   &  & & & & \\
10. & is member of cluster with   & 4 & & & \checkmark & & \\
 &most seeds/intermediates?  & & & &  & & \\
11. & fraction of seeds in cluster  & 4  &  & & \checkmark& & \\

12. & fraction of seeds and  & 4  &  & & \checkmark & & \\
& intermediates in cluster  & &  & &  & & \\
13. & query likelihood on KB text & - &  & & & content (text) & \checkmark \\

14. & in-degree of node  & - &  & & & global (KB) & \\

15. & clustering coefficient  & - &  &  & & global (KB) & \\
\bottomrule
\end{tabular}
\end{center}

\end{table}

\bigskip \noindent \textbf{Seed and intermediate features (\#1--4, Steps 1--2).} Seed and intermediate nodes are distinguished by two binary features. For all the nodes in the intermediate graph, we compute and retain their betweenness centrality and their PageRank score as features.

\bigskip \noindent \textbf{Border features (\#5--12, Steps 3--4).} We introduce a feature indicating the border nodes. We leverage information from the clustering step by associating each node with its average proximity $\bar{\sigma}(x,C_i)$ (Equation \ref{formula:score}) to the nearest cluster $C_i$. This feature is also used as an unsupervised baseline in the experimental evaluation.  Since border nodes can be associated with more than one cluster (e.g., \gist{Conservation} in Figure \ref{fig:extendedclusters}) we additionally add features capturing the sum (and average) proximity to all clusters (cf.\ Equation \ref{formula:score}).

We assume that the more seed nodes are members of a cluster, the more relevant this cluster is for expressing the gist of the image-caption pair. This assumption is expressed in two features. First, we include a binary feature indicating which nodes belong to the cluster with the highest number of seed nodes. Moreover, a second feature indicates for each node the number of seed nodes in the cluster that contains it -- e.g., if a node is a member of a cluster with two seed nodes, the value of this feature is set to 2 -- (we sum over all clusters a node belongs to, e.g., in case it is contained in multiple clusters). Exploiting the potential benefit of the joint set of seed and intermediate nodes, we similarly compute membership of the cluster with the highest number of nodes that are seed or intermediate nodes, as well as the total number of seed and intermediate nodes found in the cluster containing a node.

\bigskip \noindent \textbf{Content features (\#13).}  We include a content-based similarity measure for image-caption pairs. For this we concatenate all (distinct) entity mentions from the caption and all object annotations from the image as a keyword query. We use the query to retrieve textual content associated with article and category nodes using a query likelihood model with Dirichlet smoothing (cf. study of smoothing methods in language models ~\cite{Zhai2004}).
The retrieval model is then used to rank nodes in the candidate set relative to each other. We use this ranking as a baseline for the experimental evaluation and include the reciprocal rank as a node feature.

\bigskip \noindent \textbf{Global features (\#14--15).}  Finally, we include global node features that are independent of the image-caption pair. To this end, we compute the in-degree of the nodes in the knowledge base (i.e., the number of incoming links for a Wikipedia page or category), in order to characterize the number of their neighbors and, accordingly, their \emph{a priori} prominence in the knowledge graph. To capture the
degree in which those neighbors are found within clusters of densely interconnected nodes, we additionally compute for each node its clustering coefficient, namely the proportion of the neighbors of a node that are also neighbors of one another~\cite{Watts:1998}: this has been found in previous work \cite{SteyversT05,Griffiths07topicsin} to be for semantic networks far greater than that of a random graph, e.g., due to their topically-oriented structure.


\bigskip \noindent \textbf{Learning-to-rank model.}  Our generated feature vectors, where each vector represents a concept from our gist candidate concepts for one image-caption pair, serve as input for a list-wise learning-to-rank model~\cite{li2011letor}.  In a learning-to-rank setting, each image-caption pair is represented as a query, which at test time governs a ranking function. Given a query (i.e., an image-caption pair), each document to be ranked (i.e., a gist candidate concept from the knowledge base) is thus represented by a feature vector, and the algorithm learns from labeled data a retrieval function such that the computed ranking scores produce the best possible ranking according to some evaluation metric or loss function.

In our experiments, we use the RankLib~\footnote{\url{https://sourceforge.net/p/lemur/wiki/RankLib}} implementation of the Coordinate Ascent method -- a local search technique that iteratively optimizes a multivariate objective function by solving a series of one-dimensional searches \cite{MetzlerC07} -- and Mean Average Precision (MAP) as training metric with a linear kernel.

\section{Experiments}\label{experiments}
\noindent In the following, we investigate our proposed approach according to several aspects which are formulated in ten different research questions (RQs). Our concept ranking task is benchmarked in RQs 1 through 3, where we first evaluate our entity linking strategy to create the seed nodes (RQ1), look at the suitability of different semantic graphs (RQ2), and perform extensive feature analysis (RQ3). RQ4 and RQ5 evaluate instead different aspects related to the benefit of filtering gist candidates. The role of automatic object detection and caption generation is addressed in RQ6--10.

\medskip \noindent\textbf{Gold standard.}  To the best of our knowledge there is no public dataset providing gist annotations for image-caption pairs.\footnote{Our previous work in \cite{weiland14a} provides us only with images.} Consequently, we make use of our experiments of the dataset for understanding the message of images covering the topic of non-literal and literal image-caption pairs introduced in our previous work~\cite{weiland16a}. 


From the newspaper `The Guardian', `Our World' magazine, and the website of the `Union of Concerned Scientists'\footnote{\url{https://www.theguardian.com}; \url{https://ourworld.unu.edu}; \url{http://www.ucsusa.org}} we first manually collect image-caption pairs related to the topic of global warming, which provides us with a domain with many non-literal media icons~\cite{Perlmutter2004,Drechsel2010}. Specifically. we consider six related sub-topics:  sustainable energy, endangered places, endangered species, climate change, deforestation, and pollution. The collected image-caption pairs are non-literal pairs, thus providing us with a realistic, yet challenging dataset.
Alternative descriptive captions are then created for each image to obtain literal image-caption pairs. 
The result consists of a balanced collection of 328 image-caption pairs (164 unique images).

In order to benchmark the proposed approach for selecting and ranking the gist nodes, and to narrow down the potential of noise given by automatic object detection, we let annotators assign bounding boxes and object labels from a predefined list of concepts. 
To make our results comparable with automatic image object detectors, the list of selectable concepts is limited to objects that are depictable in the image: to this end, the annotators used a set of 43 different concepts (e.g., \word{windmill}, \word{solar panel}, \word{orangutan}) to annotate the visible objects in the images.  

We define the understanding of image-caption pairs as a concept ranking task (Section \ref{problemStatement}), where the message of an image can be represented by several concepts. To provide us with ground-truth judgements for evaluating gist selection and ranking, experts select concepts from the knowledge base (i.e., Wikipedia pages or categories) which best represent the message.
These concepts are graded by the annotators on the basis of their relevance levels, ranging from 0 (non-relevant) to 5 (most relevant). In the following we will refer to concepts with grade 5 as \textbf{core gists} and to concepts with level 4 or 5 as \textbf{relevant gists}. 
We assume that a pair can only have one concept graded with level 5: this concept represents the most relevant aspect of the gist. 
For the non-literal pairs it is often the case that the core gist corresponds to one of the before mentioned six aspects of the domain of our testbed, such as \gist{Endangered Species} (Figure~\ref{fig:nonliteralexample}). A corresponding core gist for a literal pair is \gist{Orangutan} (Figure~\ref{fig:literalexample}).

%

Our annotators produced a dataset of 8,191 gist annotations in total ($\approx$25 per image-caption pair), 3,100 of which have a grade of 4 or higher. The list of gold-standard gists in the dataset are grouped by topical gists, which can be seen as some sort of core gist.
Among all relevant nodes in this study 54,6 \% of all gist nodes are Wikipedia pages and 45,4\% are categories.

Compared to other benchmarking datasets in the field of computer vision, ours is a rather small dataset consisting of `only' a few hundred image-caption pairs. However, what we primarily annotated as gold standard is the ranking of gist nodes -- over 8,000 in total: this is what we use to evaluate the performance on concept ranking, i.e., our core task (cf.\ Section \ref{problemStatement}). As such, it compares favourably with other datasets for related tasks in the field of language and vision \cite[\emph{inter alia}]{hodosh2013imagedescription,shutova16}. Furthermore, this is the first test collection for literal and non-literal image-caption pairs with gold standard gist annotations and simulated object tags.~\footnote{\url{https://github.com/gistDetection/GistDataset}}

%


\medskip \noindent \textbf{Experimental setup.} We use a combined knowledge base aligning Wikipedia (WEX dump from 2012), Freebase (from 2012), and DBpedia (from 2014). This knowledge base is used for entity linking, deriving edges for the graph, and the content-based retrieval methods. 
As relatedness measure (Section \ref{bordergraph}), we use the metric $\sigma^{(-1)}$ from Hulpus et al.~\cite{Hulpus2015pathbased}. We use their settings for hyperparameters $\alpha=0.25$ and take the $k=3$ shortest paths.

\medskip \noindent \textbf{Evaluation metrics.} We evaluate with five-fold cross validation using standard retrieval metrics such as mean average precision (MAP), normalized discounted cumulative gain (NDCG), and Precision@k. Unless noted otherwise we binarize the assessments to relevant and non-relevant gists.

\medskip \noindent \textbf{Baselines.} 
We compare our approach with three different baselines. 
The first is a content-based method using the texts of Wikipedia article and category pages to construct a query likelihood model with Dirichlet smoothing~\cite{Zhai2004}.
For a given image-caption pair used as the query, we evaluate the resulting ranking of gist candidates according to the ranking of the probabilities given by the query likelihood model (Table~\ref{tab:graphRanking}, Wikipedia).  
The second baseline generates a ranking according to the relatedness measure computed in Step 4b. As a candidate node can be a member in several clusters, we consider the maximum relatedness score for the ranking (Table~\ref{tab:graphRanking}, max node-cluster relatedness).
A  third baseline instead randomly ranks the seed nodes (Table~\ref{tab:candidates}, Baseline Random Seeds), so as to assess the need for external knowledge.
Finally, the approach using a state-of-the-art object detector is compared with a separate baseline, which generates a ranking according to the confidence values of each detected object (Table~\ref{Tab:MS}, Baseline MS tag). 

\subsection{RQ1: Seed node linking (Step 1) -- Which strategy finds the best seed nodes?}\label{rq1}

\begin{table}[t]
\begin{center}

\scriptsize

\caption{Number of image and caption nodes after entity linking.
\label{tab:entityStats}
}
\begin{tabular}{cccccccc}
\toprule

 & \multicolumn{2}{c}{Non-Literal} && \multicolumn{2}{c}{Literal} && Overall\\
\cmidrule{2-3} \cmidrule{5-6}
 & Image & Caption && Image & Caption && \\
\midrule

Unique nodes & 43 & 674 && 43 & 298 & &806 \\
 
Total occurrences & 640 & 1612 && 640 & 894 & &3780\\
\bottomrule
\end{tabular}

\end{center}
\end{table}

\noindent We first evaluate the entity linking performance of the simple string-match method used in Step 1 to produce a set of image nodes and caption nodes. These together form the set of seed nodes. We use a separate gold standard to evaluate the correctness of the established links (i.e., not the same gold standard used in RQ2 and RQ3).  That is, in order to provide us with a ground truth to evaluate RQ1, annotators separately assessed links between entity mentions from the caption and objects of the image to nodes in the knowledge base, which were validated for correctness.


\medskip \noindent \textbf{Image and caption nodes.}
Table~\ref{tab:entityStats} shows that a total of 806 different Wikipedia concepts (i.e., pages or categories) are linked across all pairs of images and captions for a total of 3,780 links. Overall, only five noun phrases in captions could not be linked to the knowledge base (e.g., \word{underwater view}). 
%
Images make use of an object vocabulary of 43 different nodes with a total of 640 links across all images (since each image is manually paired with a literal and non-literal caption, there is no difference between the columns).
We observe a much wider range of nodes when linking entity mentions in the caption. In particular we notice a smaller vocabulary for literal image-caption pairs (298 unique nodes) compared to non-literal pairs (674 unique nodes), where each concept is mentioned about three times on average. However, we find that the caption nodes from literal versus non-literal pairs nearly have no overlap.

\medskip \noindent \textbf{Entity linking.}
The set of seed nodes is given by the union of image and caption nodes. In Step 1 we link object labels and entity mentions to article nodes (String2Article). However, the same procedure could have been applied to category names as well (String2Category). We first compare these two methods to entity links produced by TagMe, a state-of-the-art system \cite{ferragina2010tagme}. Furthermore, we use the retrieval index of texts associated with nodes and output the top ranked node (Wikipedia index).
%
%
Table \ref{tab:entityLinking} presents precision and recall achieved by these four methods on the set of all 806 unique image/caption nodes. We find that all methods perform reasonably well, where the category-based linking strategy cannot associate a vast majority of 581 objects / mentions. In particular, we find that our heuristics in Step 1 outperforms TagMe and is better in precision than retrieving from the Wikipedia index.

\begin{table}[tb]
\centering

\caption{Correctness of different entity linking methods for image and caption nodes. 
\label{tab:entityLinking}
}
\scriptsize{
\begin{tabular}{lll}
\toprule
Linking Method& {P} & {R} \\
\midrule
String2Article & 0.9 & 0.97  \\ 

String2Category & 1.0 & 0.27   \\ 

TagMe & 0.7 & 0.83  \\ 

Wikipedia index & 0.81 & 0.98  \\  
\bottomrule
\end{tabular}
}
%
%
%
%
\end{table}

\medskip \noindent \textbf{Discussion.} The TagMe system poorly performs on our dataset despite being a strong state-of-the-art entity linking system. Manual inspection revealed that TagMe is particularly strong whenever interpretation and association is required, for instance to disambiguate ambiguous names of people, organizations and abbreviations. In contrast, the concepts we are linking in this domain are mostly common nouns, for which Wikipedia editors have done the work for us already. In the remaining cases that need disambiguation, our heuristic is likely to encounter a disambiguation page. At this point, we are using a well-known disambiguation heuristic by using graph connections to unambiguous contextual mentions/objects. We conclude that our simple entity linking method on articles works much better than on categories and as well as TagMe.

\medskip \noindent \textbf{Summary of findings.} We propose to use objects that have been manually extracted from the image and entity mentions from the caption of the pair, and apply a simple string-matching strategy for linking those objects and entity mentions onto nodes, which we call the seed nodes, without direct disambiguation. The actual disambiguation is then implicitly achieved in the subsequent steps of graph traversal and re-ranking. We show that this straightforward ``lazy'' linking strategy provides comparable results to state-of-the-art algorithms.

\subsection{RQ2: Distribution of relevant gist nodes (Steps 2--4) -- Which graph is best?}


\begin{table}[tb]
\centering {
\scriptsize
\caption{Quality of gist candidate selection method. Significance is indicated by * (paired t-test, p-value $\leq0.05$).
\label{tab:binGist}
}
\begin{tabular}{lccccc}
\toprule
   & Avg cands.\ & P &  R & F1 & $\Delta$F1\%\\
\midrule
Seeds & 8.6 & 0.18 & 0.16 &  0.17 &  0.0 \\

Intermediates &  11.4 & 0.19 & 0.22 & 0.21 & +19.0\%* \\

Top Borders &  31 & 0.09 & 0.30 & 0.14 & -21.4\%*  \\
\bottomrule
\end{tabular}
}
\end{table}

\medskip \noindent \textbf{Benefits of graph expansion.} We first investigate whether good gist nodes are found in close proximity to the depicted and mentioned seed nodes. To this end, we distinguish proximity in the three expansion layers of seed, intermediate, and top-k border nodes (Step 1, 2, and 4b, respectively), and evaluate the benefits of each graph expansion by studying how precision and recall change with respect to the selection of relevant gist nodes for each expansion step. The results are presented in Table~\ref{tab:binGist}, where we provide precision, recall, and balanced F-measure, together with the number of average candidates per image-caption pair. In order to judge the significance of improvement for F1 we evaluate the relative increase in precision, on a per-image-caption-pair basis, and report the average (denoted $\Delta$). Significance is verified with a paired-t-test with level 0.05.

We find that especially the expansion into the intermediate graph increases both recall and precision. While the increase in F1 is relatively small, it is statistically significant across the image-caption pairs, where it yields an average increase of 19\%. The expansion into the border graph of Step 3 and its contraction to the closest border nodes in Step 4b yields the new set of top border nodes. While it increases recall quite drastically, the loss in precision leads to a significant loss in F1 (over the seed set).



\medskip \noindent \textbf{Distribution of high-quality gists.} We next change perspective and ask in which expansion set the majority of high-quality gists are found. Initially, we hypothesized that especially for non-literal caption pairs, fewer good gists will be found in the seed set, which motivated the graph expansion approach. Accordingly, we separately report findings on literal and non-literal subsets. We study two relevance thresholds in Table \ref{tab:distribution}, for relevant gists (grade 4 or 5) as well as a stricter threshold including the core gists (grade 5 only).  

Focusing on the distribution of relevant gists,  we notice that more than half of the gists are already contained in the seed set, and about 20\% are found in the intermediate set. The much larger border set still contains a significant portion of relevant gists. Focusing on the differences between literal and non-literal pairs, we find that there are no significant differences between the distributions. Where gists with grade 4 or 5 are highly relevant, they still include the most important visible concepts for non-literal image-caption pairs.
However, regarding the distribution of gists with grade 5, we notice that 71\% of high-quality gists in literal pairs are found in the seed set, which is in contrast to only 58\% for non-literal pairs. Also, for non-literal image-caption pairs we found the most useful gists in the set of border nodes with high cluster proximity.

\medskip \noindent \textbf{Discussion.} We confirm that many relevant and high-quality (grades 4 and 5) gists are found in the seed set and the node neighborhood. The large fraction of nodes available in the border set (compared to the intermediate set) suggests that limiting the intermediate graph expansion in Step 2, to be between seed nodes is too restrictive. We see our initial assumption confirmed in that literal image-caption pairs, which is where most of the related work is focusing on, contain more visible gists, and those are directly visible/mentioned. For non-literal pairs, the high-quality gists are not only invisible, but also more often only implicitly given. Nevertheless, the graph-based relatedness measures are able to identify a reasonable candidate set.

\begin{table}[tb]
\centering{
\scriptsize 
\caption{Statistics about proportion of relevant (grade 4 and 5) and core gists (grade 5).
\label{tab:distribution}
}
\begin{tabular}{lcccccc}
\toprule
 & \multicolumn{3}{c}{Grade 4 or 5} & & \multicolumn{2}{c}{Grade 5} \\ 
\cmidrule{2-4}  \cmidrule{6-7}
 & All & Non-Lit.\ & Literal & & Non-Lit.\ & Literal \\
\midrule 
Seeds & 53.79\%  & 53.46\% & 53.96\%& & 57.89\% & 70.75\%\\

Intermediates & 21.05\% & 21.70\% & 20.73\%  && 07.89\%& 17.92\%\\

Borders & 25.16\% & 24.84\% & 25.32\% & &34.21\%& 11.32\%\\
\bottomrule
\end{tabular}
}
\end{table}

\medskip \noindent \textbf{Summary of findings.}  We study the distribution of highly relevant gist nodes and whether good gist nodes are found in close proximity to the depicted and mentioned seed nodes. We distinguish proximity in the three expansion layers of seed, intermediate, and top-k border nodes and evaluate the benefits of each graph expansion. We show that while the gist nodes for about half of the studied image-caption pairs are among the seed nodes, for the other half one must look for the gist further away from the seeds, especially for non-literal pairs.




\subsection{RQ3: Learning to rank image gists (Step 4--5) -- Which features reveal the gist nodes?}

\begin{table*}[t]
\centering
\scriptsize
\caption{Entity ranking results (grade 4 or 5) of supervised learning-to-rank. Significance is indicated
by {*} (paired t-test, p-value $\leq0.05$). \label{tab:graphRanking} }
\begin{tabular}{lllllllllllll}
\toprule
 & \multicolumn{4}{l}{Both } & \multicolumn{4}{l}{Non-Literal } & \multicolumn{4}{l}{Literal}\\
\cmidrule{2-13} 
 & \tiny{}MAP  & $\Delta$\% & \tiny{}NDCG & \tiny{}P   & \tiny{}MAP & $\Delta$\%  & \tiny{}NDCG  & \tiny{}P   & \tiny{}MAP & $\Delta$\% & \tiny{}NDCG  & \tiny{}P   \\
& &  & @10  & @10  &  &  & @10 & @10 &  &  & @10 & @10  \\
\midrule
{All }  & 0.69 & 0.0  & 0.73  & 0.7  & 0.56 & 0.0 & 0.6  & 0.56    & 0.82  & 0.0 & 0.87  & 0.84 \\
Features & \\
{All But}  & 0.66 & -4.4{*} & 0.7  & 0.67    & 0.54  & -6.9{*}  & 0.57  & 0.55 & 0.78 & -4.9{*} & 0.83  & 0.8  \\
 Borders &\\ 
{All But}  & 0.69 & -0.3 & 0.71  & 0.7   & 0.56 & +0.7 & 0.57   & 0.57   & 0.81& -0.1   & 0.85 & 0.83 \\
 Interm. & \\
{Only }  & 0.63 & -8.7{*} & 0.64  & 0.64    & 0.52   & -10{*}& 0.54  & 0.52   & 0.73 & -11{*} & 0.74  & 0.76 \\
Borders&\\
\midrule

\multicolumn{13}{c}{Baselines} \\
\midrule
Wikipedia& 0.43 & -38{*} & 0.48  & 0.37& 0.43 & -24* & 0.46& 0.37 & 0.44 & -46* & 0.37 & 0.49 \\
Max node-& 0.27  & -57{*} & 0.57  & 0.30  & 0.24 & -57* & 0.59 & 0.31 & 0.31& -62*& 0.31& 0.54 \\
cluster&  \\
\bottomrule
\end{tabular}

\end{table*}


\noindent We next evaluate the overall quality of our supervised node ranking solution (Section \ref{sec:step5}). We further inspect the question of whether features generated by global and local graph centrality measures, especially those derived from border graph expansions, enhance the overall gist node ranking.  Moreover, we use our supervised learning-to-rank approach to evaluate the benefit of the feature sets collected over the various steps of our pipeline. To this end, we train a learning-to-rank model using the ground-truth judgements of relevant gist nodes (cf.\ the previous description of our gold standard). Due to the limited amount of image-caption pairs, we opt for a 5-fold cross validation using each image-caption pair as one ``query'': in this way we are able to predict 328 node rankings for all image-caption pairs, while keeping training and test data separate.

We study the research question with respect to both, non-literal, and literal pairs and report ranking quality in terms of mean-average precision (MAP), NDCG@10, and precision (P@10) of the top ten ranks. We train and compare four models based on our feature set (Table~\ref{tab:features}): (i) all features, (ii) all features except for the border features, (iii) all features except for the intermediate features, (iv) the subset of border features only.  This helps us understand and assess the different aspects of content and graph-based semantic relatedness. Moreover, we implemented two baselines (using the features highlighted in Table \ref{tab:features}):  One retrieves Wikipedia text using the query likelihood model on all entity mentions and object annotations concatenated, the other is based on an unsupervised ranking according to the maximal node-cluster relatedness measure $\sigma$ described in Step 4.
The results, presented in Table~\ref{tab:graphRanking}, are tested for significance (p-value $\leq0.05$).  

\medskip \noindent \textbf{Overall results.}  Our approach achieves relative high ranking performance of 0.69 MAP across all image-caption pairs. As expected, ranking non-literal image-caption pairs is much harder (MAP: 0.56) than for literal pairs (MAP: 0.82). Yet, even in the non-literal case, more than half of the nodes in the top 10 are relevant. 
%
Thanks to our approach, we are able to beat the baselines by a large margin. The baseline which ranks nodes by the query likelihood model on all entity mentions and objects achieves a MAP of 0.43 (being 38\% worse). The baseline which just includes the max node-cluster relatedness obtains an even worse performance of 0.27 for MAP, even though both achieve the same P@10 performance.

\medskip \noindent \textbf{Feature analysis.} We next look at the contribution of different types of features (cf.\ Table~\ref{tab:features}), and compare the performance changes in an ablation study (cf.\ Table~\ref{tab:graphRanking}).
When only the border features are used the ranking quality drops significantly, by up to $11\%$. 
This indicates the importance of the global graph, intermediate graph, and content-based features.
The maximum quality drop is for literal pairs, which indicates that the literal pairs benefit less from the graph expansion to two hops around seeds and intermediates than the non-literal pairs. 
When we use all the features except the border ones, the performance drops by up to $7\%$. This drop is stronger for non-literal pairs, reinforcing the fact that non-literal pairs benefit more from the border features.   
This performance drop cannot be detected when the intermediate features are not considered (not significant), the results are more or less comparable to the results of the complete feature set. 

\medskip \noindent \textbf{Summary of findings.} We thoroughly analyze global and local graph features, as well as content features for image gist ranking using a learning-to-rank approach. We show that the combination of the two types of features achieves the highest accuracy. Furthermore, we show the superiority of our solution in comparison with both supervised and unsupervised baselines. The fact that our full re-ranking pipeline improves so drastically over both a retrieval and a cluster-relatedness baseline demonstrates the benefit of our approach. 



\subsection{RQ4: Ranking different sets of candidate gists -- Which node types reveal the gist?}

\noindent The statistics from Table~\ref{tab:distribution} indicate that gist nodes are scattered across all sets of concepts gathered throughout our pipeline (i.e., seeds, intermediate and border nodes). Consequently, we next investigate the ability of our supervised model in detecting gists across these different regions: we benchmark this by conducting an ablation study and comparing different sets of candidate gists as input, which are collected from the different regions of our semantic graphs. We evaluate the performance of our learning-to-rank approach on four different node sets: (i) seed nodes, (ii) seed and border nodes, (iii) seed and intermediate nodes, and (iv) all three node types -- across all combinations we only consider the top-k nodes (k = 20). We compare this against a baseline that uses a random subset of the seed nodes.


\begin{table*}[]
\scriptsize
\caption{Evaluation of different candidate sets, abbreviated as seeds (S), intermediate (I), and border (B) nodes. Entity ranking results (grade 4 or 5) of supervised learning-to-rank. Significance is indicated by * (paired t-test, p-value $\leq0.05$). \label{tab:candidates}}
\begin{tabular}{lllllllllllll}
\toprule
\multicolumn{1}{l}{}& \multicolumn{4}{l}{Both} & \multicolumn{4}{l}{Non-Literal} & \multicolumn{4}{l}{Literal} 
\\
\cmidrule{2-13} 
\multicolumn{1}{l}{}&\tiny{} MAP & $\Delta$\% &\tiny{} NDCG &\tiny{} P & \tiny{} MAP &$\Delta$\% & \tiny{}NDCG &\tiny{} P & \tiny{} MAP & $\Delta$\% &\tiny{} NDCG &\tiny{} P \\
\midrule
\multicolumn{13}{c}{Top 20}\\
\midrule
S, I \& B & 0.69 & 0.00 & 0.73 & 0.7 & 0.56 & 0.00 & 0.6 & 0.56 & 0.82 & 0.00 & 0.87 & 0.84\\
S, I & 0.57 & -17* & 0.71 & 0.68 & 0.46 & -18* & 0.58 & 0.54 & 0.67 & -16* & 0.83 & 0.81\\
S, B & 0.48 & -30* & 0.65 & 0.58 & 0.31 & -45* & 0.47 & 0.38 & 0.64 & -20* & 0.83 & 0.78\\
S & 0.31 & -54* & 0.61 & 0.52 & 0.21 & -63* & 0.43 & 0.33 & 0.42 & -48* & 0.80 & 0.72\\
\midrule
\multicolumn{13}{c}{All}\\
\midrule
S, I \& B & 0.56 & -18* & 0.62 & 0.61 & 0.43 & -23* & 0.50 & 0.50 & 0.68 & -15* & 0.74 & 0.72  \\
\midrule
\multicolumn{13}{c}{Baseline}\\
\midrule
\tiny{}{Random Seeds} & 0.17 & -75* & 0.41& 0.35& 0.14 & -75* & 0.35&0.26 & 0.2 & -76*& 0.49&0.35 \\
\bottomrule
\end{tabular}
\end{table*}

The results, shown in Table~\ref{tab:candidates}, indicate that the best MAP scores can be achieved with the complete set of candidate nodes (S, I \& B MAP: 0.68), that is, by providing candidate gists as found among all seed, intermediate and border nodes. This observation holds for both, non-literal, and literal pairs. 
Throughout both, non-literal, and literal pairs, the candidate set provided by seed and intermediate nodes performs better than the one provided by seed and border nodes. 
An additional interesting aspect is the performance comparison of the seed nodes with respect to the literal and non-literal pairs, where the MAP for the non-literal pairs is half (MAP: 0.42 vs.\ 0.21). 

\medskip \noindent \textbf{Summary of findings.} We investigate which node types help reveal the gists, and evaluate the performance on four different node sets. We show that the best results are achieved with the complete node set. Furthermore, we corroborate our previous finding that it is harder to detect the gist of non-literal image-captions than literal ones. This effect can especially be observed for the seed-only candidate set: even though the amount of relevant gists for non-literal and literal pairs are nearly equal, the non-literal pairs have less core gist within the seeds (cf.\ Table~\ref{tab:distribution}), which directly influences the quality of the ranking. This is because the gist of non-literal pairs cannot be found explicitly among the entity mentions.

\subsection{RQ5: Filtering candidate gists -- Do related concepts better reveal the gist?}


\noindent Although results for RQ4 show that the best results can be achieved by considering all node types as input to the supervised model, we are still left with the question of whether some candidates are better than others. Consequently, we next look at whether considering only the top-k nodes from the candidate set of seed, intermediate, and border nodes helps improve the results -- i.e., by assuming that there are only a few nodes related to the initial query (seed nodes). We propose to use our relatedness measure \cite{hulpus2013} as an indicator to select the top-k most relevant nodes and potentially filter out distracting ones.

In Table~\ref{tab:candidates} we compare the performance using the complete set of candidate nodes (S, I \& B, line 5) with a subset obtained by selecting its top-20 elements (line 1). The performance loss of nearly 20\% for both type of pairs (MAP: 0.56) indicates the usefulness of using a relatedness measure as a pre-filtering step. The non-literal pairs benefit more from the relatedness-based selection than the literal pairs ($\Delta$\%: -23 vs. -15), arguably the hardest subset of data.

\medskip \noindent \textbf{Summary of findings.} We propose to identify the gists by ranking only the top-k candidates, obtained using a relatedness measure: the results indicate that relatedness-based filtering helps for both image-caption pair types, literal and non-literal.

\subsection{RQ6: Finding relevant gist types -- Are image gists depictable concepts?} 
\noindent One of the main objectives of our work is to develop a framework to identify the message (gist) conveyed by images and their captions, when used either literally or non-literally (Section \ref{problemStatement}). Consequently, we next investigate the type of concepts that humans find suitable as gists, that is, whether the gist concepts as selected by annotators tend to be depictable or non-depictable. Note that here, we are not looking for the visibility of  concepts in a specific image \cite{Dodge}, but rather investigate whether the message of the image-caption pair can in general be depicted. Our hypothesis is that the problem of gist detection is particularly challenging for image-caption pairs whose gist is a concept that is not depictable. 

For a subset of the gold standard pairs, annotators labeled each relevant gist concept as \textit{depictable, non-depictable, or undecided}.  On average the fraction of depictable core gists is $88\%$ for literal pairs versus only $39\%$ for the non-literal pairs.
On the larger set of all relevant gists, $83\%$ are depictable for literal pairs versus $40\%$ for the non-literal pairs. The annotation task, in practice, tends to be rather difficult for humans themselves, as reflected in an inter-annotator agreement (Fleiss' kappa~\cite{fleiss1971mns}) of $\kappa=0.42$ for core gists and  $\kappa=0.73$ for relevant gists.

\medskip \noindent \textbf{Summary of findings.}  We study whether gists are in principle depictable or not. The results of our annotation study are in line with our initial assumption that literal pairs tend to have depictable concepts as gist, whereas the message of non-literal pairs is conveyed through a predominant amount of non-depictable concepts. Generally, this indicates that the core message of images does not necessarily correspond to objects that are depicted, i.e., explicitly to be found within the image: as such, it motivates semantic approaches like ours that aim at going beyond what is found explicitly in the image and accompanying text, to detect the \emph{purpose} for which an image is used.



\begin{table}[t]
\centering

\caption{Ranking results (grade 4 or 5) according to different input signal and their combination (automatically generated and single signals). `M' and `A' indicate manually and automatically produced object labels and caption text, respectively. Significance is indicated by * (paired t-test, p-value $\leq0.05$).}
\label{Tab:MS}
\tiny
\begin{tabular}{lllllllllllllll}
\toprule
& & & \multicolumn{4}{l}{Both } & \multicolumn{4}{l}{Non-Literal } & \multicolumn{4}{l}{Literal}\\
\cmidrule{4-15} 
\# & object & caption &  MAP  & $\Delta$\% & NDCG & P   & MAP & $\Delta$\%  & NDCG  & P   & MAP & $\Delta$\% & NDCG  & P   \\
&labels&text& &  & @10  & @10  &  &  & @10 & @10 &  &  & @10 & @10  \\

\midrule
\multicolumn{15}{c}{{Image $+$ caption}} \\
\midrule
1 &  M &  M &0.74&0.00&0.71&0.71&0.64&0.00&0.59&0.58&0.83&0.00& 0.84&0.84\\
2 &  M &  A  &0.48&-35* & 0.63 & 0.56 & 0.36 & -44* &0.45 &0.38 &0.61 & -27* &0.80 &0.73\\
3 &  A &  M  &0.43& -42* &0.58 & 0.53 & 0.40 & -38* &0.49 & 0.44 &0.46 &-45* & 0.68 &0.61\\
4 &  A &  A &0.14& -81* & 0.28 & 0.23 &0.09 &-86* & 0.17 & 0.14 &0.20 & -76* &0.39 &0.32\\

\midrule
\multicolumn{15}{c}{{Image only}} \\
\midrule
5 &  M &  -- &0.48& -37* &0.65 &0.57 & 0.28 &-47*&0.40 &0.33 &0.68 & -28* &0.89 & 0.82\\
6 &  A &  --  & 0.13 & -84* & 0.24 &0.20 & 0.06 &-90* & 0.13 &0.11 &0.20 &-80* &0.35 &0.29\\
\midrule
\multicolumn{15}{c}{{Caption only}} \\
\midrule
7 &  -- &  M &0.38& -50* & 0.54 & 0.49 &0.31 & -52* & 0.40 & 0.35 & 0.45 &-48* &0.67 &0.63\\
8 &  -- &  A &0.07& -92* & 0.15 &0.12 & 0.05 &-93*&0.10 &0.08 &0.09 &-89* &0.19 &0.16 \\
\midrule
\multicolumn{15}{c}{{Baseline}} \tabularnewline
\midrule
9 & -- & -- &0.02 &-97* &0.26 & 0.05&0.01 &-98* &0.15 & 0.03&0.03 & -98*& 0.36& 0.07\\
\bottomrule
\end{tabular}

\end{table}

\subsection{RQ7: Manual vs.\ automatic object detection -- Do we need manual object labeling?}
\noindent All experiments carried out so far relied on a gold standard where human annotators manually assigned bounding boxes and object labels to image objects in our dataset. Consequently, we now investigate the performance of our system when the objects in the image are automatically detected with state-of-the-art image annotation tools. We make use of the Computer Vision API~\footnote{\url{https://www.microsoft.com/cognitive-services/en-us/computer-vision-api}} from Microsoft Cognitive Services \cite{fangCVPR15} -- a Web service that provides a list of detected objects and is also capable of generating a descriptive caption of the image. This experiment provides an evaluation of our method in a realistic, end-to-end setting, where images are given with accompanying captions but without manually labeled image object tags.

We first compare the tagging output of the automatic versus manual object labeling. The manual gold standard is based on a vocabulary of 43 different object labels used to annotate 640 instances over the complete dataset. The automatically labeled data amount to 171 unique object labels used to tag 957 instances. There are 131 overlapping instances between manual and automatic tags, which amounts to less than one shared tag per image, and $20\%$ overlap over the complete dataset.

We next compare the performance on our concept ranking gold standard (Table~\ref{Tab:MS}, lines 1 vs.\ 3). Although a higher performance is achieved with manual tags (MAP: 0.74), the automatic approach achieves a reasonable quality as well (MAP: 0.43). Thus, our experiments show that while there is a certain quality loss in the output predictions, our approach is stable enough to provide useful gists even when applied to the more noisy output of an automatic image annotation system. 

\medskip \noindent \textbf{Summary of findings.}  The overlap between automatic and manual image tags is rather low (20\%), and the detected objects are not always correct (e.g., a polar bear is detected as a herd of sheep). However, the automatic tags in combination with the human captions lead to a mild drop in performance on gist detection, thus indicating the viability of our approach in an end-to-end setting.

\subsection{RQ8: Manual vs.\ automatic caption generation -- Do we need human captions?}

\noindent The state-of-the-art image understanding system provided by Microsoft's Computer Vision API is able not only to tag images, but also to generate descriptions of the content of the images. As such, it provides us with a high-performing system to generate image captions \cite{fangCVPR15}. Consequently, we next investigate a research question complementary to the previous one, namely how the performance of our method is affected when the caption is automatically generated, as opposed to having been manually produced.

Similarly to RQ7, we first compare the tagging output of the automatic versus manual captions. With respect to Entity linking (Step 1, Section \ref{approach}), the manually created captions result in around 300 and 700 different entities (seed nodes) for the literal and non-literal pairs (Table \ref{tab:entityStats}). When using the automatically generated captions, these numbers shrink to 130 different detected entities only -- thus indicating that the automatic captions are less heterogeneous in meaning than the manual ones. Arguably, this is due to the fact that automatic detectors are trained to produce literal captions. To evaluate the suitability of automatic captions for gist detection, we pair manual or automatic captions with the manual image tags and provide them as input for our pipeline (Figure \ref{fig:pipeline}). The results, shown in Table~\ref{Tab:MS} (lines 1 and 2), indicate that similar to the case of automatic image labeling, our approach suffers from a mild yet clear decrease in performance (MAP: 0.74 vs.\ 0.48). Finally, we test the performance of the system when using \emph{both} automatic object labels and captions (Table~\ref{Tab:MS}, line 4): in this case, the dramatic performance decrease (MAP: 0.14) indicates that our method is robust whenever it is provided with at least one signal source (i.e., visual or textual) that is manually produced and cannot cope with purely automatically generated input.
 
\medskip \noindent \textbf{Summary of finding.} The overlap between the entities found within automatic and manual captions is low (3-10\%). The automatic captions are often short, and the focus of the captions does not always match the focus of the manual caption (e.g., the example in Figure~\ref{fig:literalexample} receives the caption "There is a bench", without considering the orangutan, although it was detected as a monkey by the automatic image tagging). The results on gist detection, however, are similar to those obtained using automatic image tagging, but drastically drop when providing the system with a purely automatically generated input (i.e., automatically generated image labels and captions).

\subsection{RQ9: Manual vs.\ automatic input -- Does an automatic approach capture more literal or non-literal aspects?}

\noindent In the previous two RQs we benchmarked the performance degradation of a manual versus automatically generated input, i.e., image labels (RQ7) and captions, (RQ8). We now turn to the complementary question of which kind of image-caption pairs are better captured by an automatic approach. That is, we investigate the question: is the output of a state-of-art image tagger and caption generator better suited to identify the gist of literal or non-literal image-caption pairs? Again, we rely on the Computer Vision API of Microsoft for such purpose. As shown in Table~\ref{Tab:MS} (line 4) using both automatic image tags and captions leads to a moderate ranking for the literal pairs (MAP: 0.20), whereas performance for the non-literal pairs is much lower (MAP: 0.09). This effect is likely due to the fact that the image understanding system we use is trained on a much different kind of data and with a purpose other than detecting (possibly, abstract) image gists. Microsoft Cognitive Service uses, in fact, a network pre-trained on ImageNet and includes a CNN, which assigns labels to image regions trained on Microsoft COCO data. Similarly, the language generation uses a language model built using 400,000 (literal) image descriptions.

The `realistic' approach (i.e., the one using human captions and automatic image labels) has a consistent performance decrease of less than 40\% for both image-caption pair types (Table~\ref{Tab:MS}, line 3 -- MAP: 0.40 and 0.46 for non-literal and literal pairs, respectively). Substituting only the manual captions with the automatic ones (Table~\ref{Tab:MS}, line 2), instead, results in a lower performance drop than when using automatic image tags for the literal pairs (MAP: 0.61 vs.\ 0.46), but a lower overall performance for the non-literal pairs (MAP: 0.36 vs.\ 0.40). Again, this is likely to be due to the caption generator being able to leverage background knowledge for literal, i.e., descriptive caption generation on the basis of the underlying language model. Such an approach, however, cannot, and is not meant to generate non-literal, topically abstract captions.

\medskip \noindent \textbf{Summary of findings.} The evaluation results across all input signal combinations confirm that gists of non-literal pairs are generally more difficult to detect. Automatic approaches can account for descriptive pairs by detecting important objects in the image and describe those in the caption. However, the automatic approaches are not able to produce high-level, abstract image descriptions that are salient to detect the gist of non-literal pairs. That is, to detect the gist of non-literal image-caption pairs, to date, we need to rely on manually produced captions, a requirement that can be dropped to detect the message of literal pairs only.  

\subsection{RQ10: Visual vs.\  textual information -- Does the image or the caption convey the gist?}

\noindent In our last RQ, we look at the role of different kinds of signals within our approach. To this end, we test the performance on gist detection when using only visual (cf. Table~\ref{Tab:MS}, lines 5--6) or textual (lines 7--8) information separately. For each modality, i.e., visual or textual, we additionally benchmark performance as obtained when using automatically versus manually created image labels or captions. That is, we additionally cast RQs 8 and 9 in a single modality setting.

Given the manual image tags as input signal only, gist detection on literal pairs suffers from a lower performance drop as when compared to non-literal pairs (cf. Table~\ref{Tab:MS}, line 5, Literal MAP: 0.68 and Non-Literal MAP: 0.28, respectively).
Using automatic object labels only additionally lowers performance, with a massive drop for non-literal gists (cf. Table~\ref{Tab:MS}, line 6, Literal MAP: 0.20 and Non-Literal MAP: 0.06). These very same trends are shown also when using either manually (line 7) or automatically (line 8) generated captions only: using textual information only lead to a high performance decrease for both image-caption pair types (Literal MAP: 0.45 and Non-Literal MAP: 0.31), which is even higher in the case of automatically generated captions (Literal MAP: 0.09 and Non-Literal MAP: 0.05). Nevertheless, all configurations are able to outperform a baseline obtained by using the Vision API's confidence values for each image directly to establish the ranking (line 9). When investigating different signal sources separately, we are able to corroborate our previous findings that the gists of literal pairs are easier to detect than the gist of non-literal ones. Besides,  given that performance substantially decreases when using only the image tags or captions, we show that image and caption are complementary sources of information to effectively detect the message of image-caption pairs. This is in line with many previous contributions from the field of multi-modal modeling that have demonstrated improvements by combining textual and visual signals.

\medskip \noindent \textbf{Summary of findings.}  The evaluation results across different modalities indicate the complementarity nature of visual and textual information for detecting the gist of both, literal and non-literal image-caption pairs. That is, by showing that performance on gist detection is reduced when the image tags or only the caption are provided, we show that both image and caption are required in order to capture the message of images.

\section{Conclusion}\label{concl}

%
%
\noindent In this paper, we presented a knowledge-rich approach to discover the message conveyed by image-caption pairs. We focused on an heterogenous dataset of literal image-caption pairs (whose topic is described through objects and concepts found in either the picture or the accompanying text), as well as non-literal ones (i.e., referring to abstract topics, such media-iconic elements found in news articles). Using a manually labeled dataset of literal and non-literal image-caption pairs, we casted the problem of gist detection as a ranking task over the set of concepts provided by an external knowledge base. Specifically, we approached the problem using a pipeline that: i) links detected object labels in the image and entity mentions in the caption to nodes of the knowledge base; ii) builds a semantic graph out of these `seed' concepts; iii) applies a series graph expansion and clustering steps of the original semantic graph to include additional, non-depictable concepts and topics within the semantic representation; iv) combines several graph-based and text-based features into a node ranking model that pinpoints the gist nodes.

%
%
Our experiments show that the candidate selection and ranking of gist concepts is a more difficult problem for  non-literal image-caption pairs than for literal image-caption pairs. Nevertheless, we demonstrated that using features and concepts from both modalities (image and caption) improves the performance for all types of pairs, a finding which is in line with research on multimodal approaches for other related tasks. Additionally, a feature ablation study shows the complementarity nature and usefulness of different types of features, which are collected from different kinds of semantic graphs of increasing richness. Finally, we experimented with a state-of-the-art image object detector and caption generator to evaluate the performance of an end-to-end solution for our task. 

The results indicate that using state-of-the-art open-domain image understanding provides us with an input that is good enough to detect gists of image-caption pairs, with nearly half of predicted gists being relevant.  However, it also demonstrates that improved object detectors could avoid a drop of 38\% mean-average precision. Additionally, the caption contains useful hints especially for non-literal pairs. However, without considering the information of the image leads to significant performance degradation.

%
%
Gist image identification is a small, yet arguably crucial part of the much bigger problem of interpreting images beyond their denotation. As such, we see this study as a starting point for research on gist-oriented image search and classification, detection of themes in images, and recommending images from the web when writing new articles for news, blogs, and Wikipedia.  But even in the simple form of casting image understanding as a concept ranking problem, we see many potential benefits for a wide range of applications: with our method, for instance, large image collections, such as Wikimedia commons (more than 30 million images) could potentially be explored in a new way by annotating the contained images with (possibly abstract) concepts from Wikipedia. We leave the exploration of such high-end task that could profit from gist detection for future work.

\section*{Acknowledgement}
\noindent We gratefully acknowledge the support of NVIDIA Corporation with the donation of
the GeForce Titan X GPU used for this research.
This work is funded by the RiSC programme of the Ministry of Science, Research
and the Arts Baden-Wuerttemberg, and used computational resources offered from the bwUni-Cluster within the framework program bwHPC. 
Furthermore, this work was in part funded through the Elitepostdoc program of the BW-Stiftung and the University of New Hampshire.



%
%

\nocite{DuWZZLST17}

\bibliographystyle{elsarticle-num}
\bibliography{tomm17-short}  

\end{document}